\documentclass[%
 aip,
 floatfix,
 pop,    % Phys. Plasmas journal
 amsmath,amssymb,
 reprint,     % single-spaced, two-column [at least w/ pop journal]
]{revtex4-2}

\usepackage{graphicx}% Include figure files
\usepackage{dcolumn}% Align table columns on decimal point
\usepackage{bm}% bold math
\usepackage[percent]{overpic}
\usepackage{floatrow}
\usepackage{siunitx}
\usepackage{subcaption}
\usepackage[utf8]{inputenc}
\usepackage[T1]{fontenc}
\usepackage{mathptmx}
\usepackage[margin=0.75in]{geometry}
\usepackage{etoolbox}

\makeatletter
\def\@email#1#2{%
 \endgroup
 \patchcmd{\titleblock@produce}
  {\frontmatter@RRAPformat}
  {\frontmatter@RRAPformat{\produce@RRAP{*#1\href{mailto:#2}{#2}}}\frontmatter@RRAPformat}
  {}{}
}%
\makeatother
\begin{document}

\preprint{AIP/123-QED}

\title{Magnetized Laser-Plasma Interactions in High-Energy-Density Systems: Parallel Propagation}%[Sample title]{Sample Title:\\with Forced Linebreak}
% Force line breaks with \\
\author{E. E. Los}
% DJS you are currently at IC, LLNL gets acknowlegded for funding the work so good enough
%\altaffiliation[Also at ]{Physics Department, Imperial College London}%Lines break automatically or can be forced with \\
\affiliation{Physics Department, Imperial College London, South Kensington Campus, London SW7 2AZ}
\author{D. J. Strozzi}
\affiliation{Lawrence Livermore National Laboratory, Livermore, CA, 94551, USA}

\date{\today}% It is always \today, today,

\begin{abstract}
  We investigate parametric processes in magnetised plasmas, driven by a large-amplitude pump light wave. Our focus is on laser-plasma interactions relevant to high-energy-density (HED) systems, such as the National Ignition Facility and the Sandia MagLIF concept.  We derive dispersion relations for three-wave interactions in a multi-species plasma using Maxwell's equations, the warm-fluid momentum equation and the continuity equation. The application of an external B field causes right and left polarised light waves to propagate with differing phase velocities.  This leads to Faraday rotation of the polarisation, which can be significant in HED conditions. Raman and Brilllouin scattering have modified spectra due to the background B field, though this effect is usually small in systems of current practical interest. We identify a scattering process we call stimulated whistler scattering, where a light wave decays to an electromagnetic whistler wave ($\omega \lesssim \omega_{ce}$) and a Langmuir wave.  This only occurs in the presence of an external B field, which is required for the whistler wave to exist. We compute the scattered wavelengths for Raman, Brillouin, and whistler scattering.
\end{abstract}

\maketitle

\section{\label{sec:Intro}Introduction}%\protect\\ %The line
%break was forced \lowercase{via} \textbackslash\textbackslash}

Imposing a magnetic field on HED systems is a topic of much current interest.  This has several motivations, including reduced electron thermal conduction to create hotter systems (such as for x-ray sources \cite{ElijahKemp}), laboratory astrophysics \cite{WillFox}, and magnetised inertial confinement fusion (ICF) schemes. If successful, they could provide efficient, low-cost alternatives to unmagnetised, laser-driven ICF. In the most successful case, the Sandia MagLIF concept ~\cite{slutz10,Gomez14}, an external axial magnetic field is used to magnetise the deuterium-tritium (DT) gas contained within a cylindrical conducting liner. A pulsed-power machine then discharges a high current through the liner, generating a Lorentz force which causes the liner to implode. The DT fuel is pre-heated by a laser as the implosion alone is not sufficient to heat the fuel to the ignition temperature. The magnetic field is confined within the liner and hence obeys the MHD frozen in law, which states
\begin{equation}
B_z\pi r^2=c
\end{equation}
where $r$ is the radius of the cylindrical liner, $B_z$ is the axial magnetic field and $c$ is a constant. Over the course of the implosion, the magnetic field strength perpendicular to the direction of compression increases as $1/r^2$. Thus, following the implosion, the magnetic field traps fusion alpha particles and thermal electrons, insulating the target and aiding ignition.

The MagLIF scheme, as well as magnetised laser-driven ICF~\cite{Jones_1986,Lindemuth_1983}, motivate us to consider magnetised laser-plasma interactions (LPI), specifically parametric scattering processes \cite{kruer88}. Parametric coupling involves the decay of a large-amplitude or ``pump'' wave into two or more daughter waves. We focus on those involving one electromagnetic (e/m) and one electrostatic (e/s) daughter wave. In order for parametric coupling to occur, the following frequency and wave-vector matching conditions must be met:
\begin{equation}
\label{e_con}
\omega_0=\omega_1+\omega_2
\end{equation}
\begin{equation}
\label{mom_con}
\vec{k}_0=\vec{k}_1+\vec{k}_2
\end{equation}
where the subscripts 0, 1 and 2 denote the pump, scattered and plasma waves, respectively. Equations \ref{e_con} and \ref{mom_con} are required by energy and momentum conservation, respectively. Parametric processes can give rise to resonant modes which grow exponentially in the plasma and remove energy from the target~\cite{lindl04}. Additionally, light waves which are back-scattered through the optics of the experiment can cause significant damage and even be re-amplified\cite{velarde93,kirkwood14,Chapman_2019}. Finally, electron plasma waves can generate superthermal or ``hot'' electrons which can pre-heat the fuel, thereby increasing the work required to compress it~\cite{lindl98}.

Laser-driven parametric processes have been extensively researched in unmagnetised plasmas. However, the advent of experiments such as MagLIF and the possibility of magnetised experiments on the National Ignition Facility (NIF)~\cite{Perkins13,Perkins14,Moody21} necessitate re-examining the impact of a magnetic field on them, which is usually neglected.  This is not unexplored territory.  For instance, prior work studied how an external axial B field affects Raman backscattering in a hot, inhomogeneous plasma~\cite{Laham98}, and the decay of circularly polarised EM waves in cold, homogeneous plasma~\cite{Stenflo10}. Recently, excellent theoretical work on a warm-fluid model for magnetized LPI has been done by Shi~\cite{shi19}. Winjum et al.\cite{winjum} have studied SRS in a magnetized plasma with a particle-in-cell code in conditions relevant to indirect-drive ICF.  This work focuses on how the B field affects large-amplitude Langmuir waves, which can nonlinearly trap resonant electrons and modify the Landau damping.  Our work ignores nonlinearity and damping, both of which are important in real systems.

Besides modifying existing processes, a background B field gives rise to new waves, one of which is an electromagnetic "whistler" wave which has $\omega \leq \omega_{ce}$, the electron cyclotron frequency. Thus, a plethora of new parametric processes involving this wave can occur, including one which we call ``stimulated whistler scattering'', in  which the pump light wave decays to an electrostatic Langmuir wave and a whistler wave. Parametric processes involving whistlers have been known for some time.  For instance, a collection of new instabilities (mostly involving whistler waves) which include purely growing, modulational and beat-wave instabilities in hot, inhomogeneous plasmas has been explored by Forslund et al.~\cite{Forslund72}. The decay of a high-frequency whistler wave into a Bernstein wave and a low-frequency whistler wave in hot, inhomogeneous plasmas have also been investigated~\cite{Stenflo11}. Additionally, parametric decays involving three whistler waves in cold, homogeneous plasmas have been studied~\cite{Kumar11}. In magnetized fusion, parametric interactions of large-amplitude RF waves launched by external antennas, for plasma heating and current drive, have been explored since the 1970s.~\cite{Porkolab_1978}

This paper aims to present the theory of magnetized LPI in a self-contained way, for a simple enough situation where that is feasible.  Namely, we consider all wavevectors parallel to the background B field, and use warm-fluid theory with multiple ion species. The results are mostly special cases of prior ones, especially by Shi, but we hope the reader benefits from a relatively simple presentation.  We obtain a parametric dispersion relation, meaning one where the pump light wave is included in the equilibrium, in the spirit of Drake et al.~\cite{Drake74}. We then study the ``kinematics'' of magnetized three-wave interactions, based on phase-matching, in HED-relevant conditions (e.g.\ for NIF and MagLIF).  We consider magnetized modifications to stimulated Raman and Brillouin scattering, as well as stimulated whistler scattering. 

The rest of the paper is organized as follows.  In section 2, we use the warm-fluid equations to derive the parametric dispersion relation. These are then linearized and decomposed in Fourier modes. Only resonant terms satisfying phase-matching are retained. In section 3, the resulting free-wave dispersion relations in a magnetised and unmagnetised plasma are discussed, along with the Faraday rotation of light-wave polarization.  Section 4 studies the impact of the external magnetic field on stimulated Raman and Brillouin scattering in typical HED plasmas.  Stimulated whistler scattering is also explored.  Section 5 concludes and discusses future prospects.

\section{\label{sec:para_disp_rel_mag}Parametric Dispersion Relations for Magnetised Plasma Waves}

This section develops a parametric dispersion relation, meaning one where the pump is included in the equilibrium. This approach is in the spirit of the paper by Drake~\cite{Drake74} for kinetic, unmagnetised plasma waves, and also for magnetised waves~\cite{Manheimer74}. Subsequent kinetic work was done which extended the Drake approach to include a background B field~\cite{cohen87,stefan87}. While our approach does not contain new results compared to the latter, we believe it is useful to work through the details explicitly - especially in a form familiar to the unmagnetised LPI community.  The upshot of the lengthy math is Eq.\ \ref{mat_coup}, which the reader should understand in physical terms before delving into the details of this section.  Our goal is expressions for the amplitude-independent $D$'s (which give linear dispersion relations) and $\Delta$'s (which give parametric coupling).

\begin{figure}[ht!]
\centering
\caption{Geometry of the experimental setup considered throughout the paper. The pump frequency, $\omega_0$, is set by the laser. An external magnetic field, $B_{eq}\hat{z}$, is imposed parallel to the propagation direction of the laser, $\hat{k}_0$. The laser is incident from vacuum on a plasma with density, $n_e$, which varies with z. The wave vector is therefore also z dependent.\label{expment_setup}}
\includegraphics[width=8.5cm, trim={0.1cm 0.0cm 0.2cm 0.2cm}, clip]{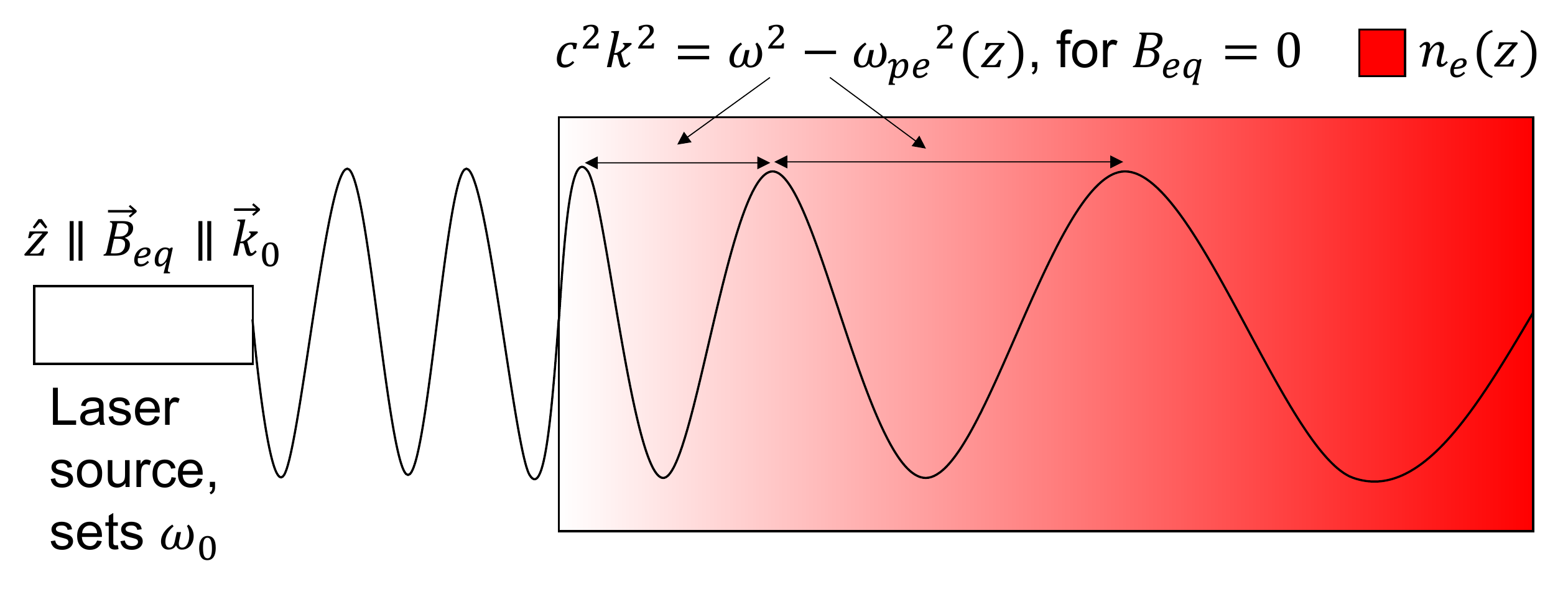}
\end{figure}
\subsection{\label{sec:gov_eqns}Governing Equations}
%Consider some vector, X:
%begin{equation}
%\vec{X}\cdot\hat{\sigma}_-=X_x-i\sigma X_y
%\end{equation}
The subscript s will be used to denote species, with mass $m_s$ and charge $q_s=Z_se$ ($e>0$ the positron charge). The subscript j will denote the wave or mode. We start with the 3D, non-relativistic Vlasov-Maxwell system with no collisions, and assume spatial variation only in the z direction. Hence, all vectors directed along $\hat{z}$ are longitudinal, and all vectors which lie in the x-y plane are transverse. An experimental configuration for which these assumptions hold is shown in figure \ref{expment_setup}. We further assume that the distribution function for species $s$, $f_s$ (particles per $dz*d^3w$, where $\vec{w}$ denotes velocity, and we have integrated over $x$ and $y$) can be written in a separable form: $f_s(t, z, \vec{w})=f_{s\perp}(t, z, \vec{w}_\perp)F_{s}(t, z, w_z)$. $f_{s\perp}$ allows for transverse electromagnetic waves, and is normed such that $\int f_{s\perp}d^2w_\perp=1$. $F_s$ is the 1D distribution (particles per $dz*dw_z$). Standard manipulations lead to the following 1D Vlasov-Maxwell system:
\begin{equation}
\label{0_ord_trns_mom_vlasov}
\partial_tF_s+w_z\partial_zF_s=-\frac{q_s}{m_s}(E_z+(\vec{v}_s\times\vec{B})_z)\partial_{w_z}F_s
\end{equation}
\begin{equation}
\label{1st_ord_trns_mom_vlasov}
(\partial_t+v_{s z} \partial_z)\vec{v}_{s\perp}=\frac{q_s}{m_s}(\vec{E}_{\perp}+(\vec{v}_s\times\vec{B})_\perp)
\end{equation}
\begin{equation}
\begin{aligned}
\label{avrg_def}
&n_s=\int F_s \mbox{d}w_z, \quad \vec{v}_{s\perp} = \int  f_{s\perp}\vec{w}_\perp\mbox{d}^2w_\perp,\\
&v_{sz} = n_s^{-1}\int F_sw_z\mbox{d}w_z
\end{aligned}
\end{equation}
\begin{equation}
\label{const_Bz}
B_z=B_{eq}=\mbox{const}
\end{equation}
\begin{equation}
\label{Max3}
\partial_t\vec{B}_{\perp}=\partial_z(E_y, -E_x)
\end{equation}
\begin{equation}
\label{Max4_trans}
\partial_t\vec{E}_\perp = c^2\partial_z(-B_y, B_x)-\frac{e}{\epsilon_0}\sum_sZ_sn_s\vec{v}_{s\perp}
\end{equation}
\begin{equation}
\label{Max4_para}
\partial_tE_z=-\frac{e}{\epsilon_0}\sum_sZ_sn_s v_{sz}
\end{equation}
$B_{eq}>0$ and the subscript eq indicates a non-zero, zeroth order background term. Poisson's equation is not listed since the inclusion of Ampere's law and charge continuity render it redundant. It is possible to satisfy Maxwell's equations (equations \ref{Max3}, \ref{Max4_trans}, and \ref{Max4_para}) by writing $\vec{E}$ and $\vec{B}$ in terms of scalar and vector potentials, $\phi$ and $\vec{A}$: $\vec{E}=-\vec{\nabla}\phi-\frac{\partial \vec{A}}{\partial t}$ and $\vec{B}=\vec{\nabla}\times\vec{A}+B_{eq}\hat{z}$.  We choose the Weyl gauge, in which $\phi=0$ and $\vec{A}=\vec{A}_\perp+A_z \vec{z}$. Faraday's law is then automatic, and the remaining Maxwell's equations become:
\begin{equation}
\label{max4_long}
\partial_t^2 A_{z}=\frac{e}{\epsilon_0}\sum_sZ_{s}n_{s}v_{sz}
\end{equation}
\begin{equation}
\label{max4_perp}
(\partial_t^2-c^2\partial_z^2)\vec{A}_{\perp}=\frac{e}{\epsilon_0}\sum_sZ_{s}n_{s}\vec{v}_{s\perp}
\end{equation}
We arrive at fluid equations by taking moments $\int w_z^p \mbox{d}w_z$ of the equation for $F_s$, for $p=$ 0, 1, and 2:
\begin{equation}
\label{0th_mom_vlasov}
\partial_t n_{s}+\partial_z(n_s v_{sz})=0
\end{equation}
\begin{equation}
\label{1st_mom_vlasov}
\partial_t(n_sv_{sz})+\partial_z\left(n_sv^2_{sz}+\frac{P_{s}}{m_s}\right)=\frac{q_sn_s}{m_s}(E_z+(\vec{v}_s\times\vec{B})_z)
\end{equation}
\begin{equation}
(\partial_t+v_{sz}\partial_z)P_s = -3P_s\partial_z v_{sz}-2\partial_zQ_s
\end{equation}
with pressure $P_s\equiv m_s \int F_s(w_z-v_{sz})^2\mbox{d}w_z$ and heat flux $Q_s\equiv (m_s/2)\int F_s (w_z-v_{sz})^3\mbox{d}w_z$. Note that the pressure is the $zz$ component of the 3D pressure tensor, \emph{not} the scalar, isotropic pressure. We can close the fluid-moment system by replacing the pressure equation with a polytrope equation of state, where $K_s$ is a constant:
\begin{equation}
\label{polytrope_eos}
P_s=n_sT_s=K_sn_s^{\gamma_s}
\end{equation}
\begin{equation}
\label{polytrope_eos_deriv_simple}
\partial_z P_s=K_s \gamma_s n_s^{\gamma_s -1} \partial_z n_s=\gamma_s T_s \partial_z n_s
\end{equation}
Common choices for linearised dynamics are isothermal $(\gamma_s = 1)$ and adiabatic $(\gamma_s = 3)$, which follows from setting $Q_s = 0$ in the pressure equation.
Let us recap the complete fluid-Maxwell system, with the substitutions  $\vec{a}=\frac{e}{m_e}\vec{A}$ (units of speed), $\omega_{ps}^2=\frac{q_s^2n_{seq}}{\epsilon_0 m_s}$, $\omega_{cs}=|\frac{q_s}{m_s}B_{eq}|$, $\mu_s=\frac{m_s}{m_e Z_s}$ and $s_s=-1 ,1$ for electrons and ions respectively:
\begin{equation}
\label{max4_long_normed}
\partial_t^2 a_{z}  - \sum_s\omega_{ps}^2\mu_s\frac{n_{s}}{n_{s eq}}v_{sz} = 0
\end{equation}
\begin{equation}
\label{max4_perp_normed}
(\partial_t^2-c^2\partial_z^2)\vec{a}_{\perp}=\sum_s\omega_{ps}^2\mu_s\frac{n_{s}}{n_{s eq}}\vec{v}_{s\perp}
\end{equation}
\begin{equation}
\label{flu_mom_long}
\begin{aligned}
&\partial_t v_{sz}+\mu_s^{-1}\partial_t a_{z} + v_{sz}\partial_z v_{sz} + \gamma_s \frac{T_s}{m_sn_s} \partial_z n_s=\mu_s^{-1}\vec{v}_{s\perp}\cdot\partial_z\vec{a}_{\perp}\\
\end{aligned}
\end{equation}
\begin{equation}
\label{flu_mom_trans}
\begin{aligned}
&\partial_t \vec{v}_{s\perp}+\mu_s^{-1}\partial_t \vec{a}_{\perp}-s_s\omega_{cs}\vec{v}_{s\perp}\times\hat{z} = -\mu_s^{-1}v_{sz}\partial_z\vec{a}_{\perp} - v_{sz}\partial_z \vec{v}_{s\perp}\\
\end{aligned}
\end{equation}
\begin{equation}
\label{mass_cont}
\partial_t n_{s}+\partial_z(n_s v_{sz})=0
\end{equation}
Terms that can give rise to parametric couplings of interest have been moved to the RHS. These involve at least one e/m wave, which will become the pump, and one e/m or e/s wave, which will become one of the daughters. All other terms have been moved to the LHS, namely those that are purely linear or contain 2nd-order terms not of interest. It is clear that the longitudinal dynamics are unaffected by $B_{eq}$ in the absence of parametric coupling, since we chose $\vec{k}||B_{eq}\hat{z}$.

\subsection{\label{sec:linearisation}Linearisation: Physical Space}
We consider parametric processes involving the decay of a fixed, finite-amplitude, electromagnetic pump to an electromagnetic and an electrostatic daughter wave, denoted by subscripts 0, 1 and 2, respectively. The daughter waves are assumed to be much lower in amplitude than the pump. We write the velocity and vector potential pertaining to each wave as an infinite sum of terms of increasing order in amplitude. We neglect all terms of second-order or higher in the pump amplitude (such as the ponderomotive term, which scales as $a_0^2$), retaining only terms which are strictly linear in wave amplitudes or involve the product of one pump and one daughter amplitude. The plasma density is approximated by the sum of a static, uniform equilibrium term, $n_{seq}$ and a perturbation induced by the electrostatic wave, $n_{s2}$. We assume that no background flows exist in the plasma ($v_{seq}=0$), no external electric fields are imposed upon it ($a_{eq}=0$), and quasi-neutrality holds ( $\sum_s q_s n_{seq}=0$). We write
\begin{equation}
\label{a_lin}
\vec{a}_{\perp}=\vec{a}_{0\perp}+\vec{a}_{1\perp}
\end{equation}
\begin{equation}
\label{a_lin_long}
a_z=a_2
\end{equation}
\begin{equation}
\label{v_lin}
\vec{v}_{s\perp}=\vec{v}_{s0\perp}+\vec{v}_{s1\perp}
\end{equation}
\begin{equation}
\label{v_lin_long}
v_{sz}=v_{s2}
\end{equation}
\begin{equation}
\label{n_lin}
n_{sz}=n_{s eq}+n_{s 2}
\end{equation}
where $\vec{a}_{j}$, $\vec{v}_{j}$ and $n_{s 2}$ are functions of $t, z$. Since we are only interested in second order terms which give rise to parametric coupling, we can linearise equation \ref{polytrope_eos_deriv_simple}:
\begin{equation}
\frac{\partial_z P_s}{n_s}=\gamma_s \frac{T_{s eq}}{n_{s eq}}\partial_z  n_{s 2}
\end{equation}
Substituting these results and equations \ref{a_lin}-\ref{n_lin} into equations \ref{max4_long_normed}-\ref{mass_cont} and keeping only coupling terms of interest, we obtain, for waves 1 and 2:
\begin{equation}
\label{max4_long_lin}
\partial_t^2 a_{2}-\sum_s\omega_{ps}^2\mu_sv_{s 2}=0
\end{equation}
\begin{equation}
\label{max4_perp_lin}
(\partial_t^2-c^2\partial_z^2)\vec{a}_{1}-\sum_s \omega_{ps}^2\mu_s\vec{v}_{s1}=\sum_s\omega_{ps}^2\mu_s\frac{n_{s 2}}{n_{s eq}}\vec{v}_{s0}
\end{equation}
\begin{equation}
\label{flu_mom_long_lin}
\begin{aligned}
&\partial_t v_{s 2}+\mu_s^{-1}\partial_t a_{2}+\gamma_s \frac{v_{Ts}^2}{n_{s eq}} \partial_z n_{s 2}\\
&=\mu_s^{-1}\left(\vec{v}_{s 0}\cdot\partial_z\vec{a}_{1}+\vec{v}_{s 1}\cdot\partial_z\vec{a}_{0}\right)
\end{aligned}
\end{equation}
\begin{equation}
\label{flu_mom_trans_lin}
\begin{aligned}
&\partial_t \vec{v}_{s 1}+\mu_s^{-1}\partial_t \vec{a}_{1}-s_s\omega_{cs}\vec{v}_{s 1}\times\hat{z}=-v_{s 2}\partial_z(\vec{v}_{s0} + \mu_s^{-1}\vec{a}_{0})\\
\end{aligned}
\end{equation}
\begin{equation}
\label{cont_eq_lin}
\partial_t  n_{s 2}+n_{s eq} \partial_z v_{s 2}=0
\end{equation}
where $v_{Ts}^2=\frac{T_{eq s}}{m_s}$. The $-v_{sz}\partial_z v_{sz}$ term in equation \ref{flu_mom_long} has been neglected because it is second order in the daughter wave amplitude. Wave 0 satisfies the same eqs. as wave 1 (i.e. Eqs. \ref{max4_perp_lin} and \ref{flu_mom_trans_lin}) without the coupling terms (RHS = 0).
For the daughter waves 1 and 2, we now have $2s+1$ scalar and $s+1$ vector equations for $2s+1$ scalar $(n_{s2}, v_{s2}$ and $a_2)$ and $s+1$ vector $(\vec{v}_{s1}$ and $\vec{a}_1)$ unknowns, with all vectors in the 2D transverse $(xy)$ plane. Our plan is to move to Fourier space, retain only linear and parametric-coupling terms, and arrive at a closed system just involving the $a$'s.

\subsection{\label{sec:Fourier_space}Fourier Decompositions}

If the variable X is used to represent the electric field, electron density or wave velocity, then X can be written as a Fourier decomposition, in which j denotes the wave (0,1,2):
\begin{equation}
\label{E_fourier}
X_{j}(t, \vec{r})=\frac{1}{2}X_{fj}e^{i\psi_j} + cc
\end{equation}
Subscript f denotes the Fourier amplitude, phase $\psi_j = (\vec{k_j}\cdot\vec{r}-\omega_j t)\equiv k_jz-\omega_j t$ and cc is an abbreviation of complex conjugate. Since all successive amplitudes will be Fourier amplitudes, the subscript f will henceforth be neglected. Wave 1 can be written in terms of two e/m waves, with either an up-shifted or a down-shifted frequency vs.\ wave 0, denoted by subscripts + and - respectively. The phase matching conditions are hence
\begin{equation}
\psi_{-}=\psi_0 - \psi_2^* \qquad \psi_{+}=\psi_0 + \psi_2
\end{equation}
Growth due to parametric coupling means the daughter-wave $k_j$ and $\omega_j$ can be complex. It is assumed that the pump amplitude is fixed (no damping or pump depletion), hence $k_0$ and $\omega_0$ are real, and $\psi_0^*=\psi_0$. We choose our definitions of $\psi_\pm$ so they and $\psi_2$ have the same imaginary part, i.e.\ the same parametric growth rate.  We also choose all frequencies to have a positive real part: the companion field for Re[$\omega]<0$ follows from the condition that the physical field is real.  Although one can mix positive and negative frequency waves, we find the analysis simpler with all Re[$\omega]>0$. Especially with magnetized waves, the discussion of circular polarization for Re[$\omega]<0$ can become confusing.

\subsubsection{\label{sec:fourier_PW}Plasma Waves in Fourier Space}

We shall eliminate $n_{s2}$ and $\vec{v}_{s2}$ in favour of the $a$'s. Substituting equation \ref{E_fourier} into equations \ref{max4_long_lin} and \ref{cont_eq_lin}, we obtain:
\begin{equation}
\label{max_long_ft}
a_{2}+\frac{1}{\omega_2^2}\sum_s\omega_{ps}^2\mu_sv_{s 2}=0
\end{equation}
and
\begin{equation}
\label{cont_simp}
n_{s2}=n_{seq}\frac{k_2}{\omega_2}v_{s2}
\end{equation}
respectively. Repeating for equation \ref{flu_mom_long_lin} gives:
\begin{equation}
\begin{aligned}
  & \left( -\frac{\omega_2}{2}v_{s2} - \frac{\mu_s^{-1}}{2}\omega_2a_2 + \frac{\gamma_sv^2_{Ts}}{2n_{eqs}}k_2n_{s2} \right) + cc \\
 &=\frac{\mu_s^{-1}}{4}PC_{s2}+cc
\end{aligned}
\end{equation}
where the parametric coupling terms are contained in $PC_{s2}$ (units of frequency*speed), and
\begin{equation}
\label{PC_def}
\begin{aligned}
PC_{s2}&= -ie^{-i\psi_2}Res_2[(\vec{v}_{s0}e^{i\psi_0})\cdot (ik_\pm\vec{a}_\pm e^{i\psi_\pm})\\
&+  (\vec{v}_{s0}e^{i\psi_0})\cdot(-ik_\pm^*\vec{a}_\pm^*e^{-i\psi_\pm^*})+(\vec{v}_{s\pm}e^{i\psi_\pm})\cdot (ik_0\vec{a}_{0}e^{i\psi_0})\\
&+(\vec{v}_{s\pm}e^{i\psi_\pm})\cdot (-ik_0^*\vec{a}_0^*e^{-i\psi_0^*}) + cc
\end{aligned}
\end{equation}
where $Res_2$ denotes terms which are resonant with mode 2. Using equation \ref{cont_simp} to substitute for $n_{s2}$:
\begin{equation}
\label{flu_mom_long_ft}
\begin{aligned}
&-\omega_2(v_{s2}+\mu_s^{-1}a_2)+\gamma_{s}\frac{k_2^2v_{Ts}^2}{\omega_2}v_{s 2}=\frac{\mu_s^{-1}}{2}PC_{s2}
\end{aligned}
\end{equation}
Rearranging for $v_{s2}$:
\begin{equation}
\label{flu_mom_long_ft_vel}
\begin{aligned}
&v_{s2} = -\frac{\omega_2P_s}{\mu_s\omega_{ps}^2}(\omega_2a_2 + PC_{s2})
\end{aligned}
\end{equation}
\begin{equation}
\label{PS_def}
P_s=\frac{\omega_{ps}^2}{\omega_2^2-\gamma_sk_2^2v_{Ts}^2}
\end{equation}
Substituting this result into equation \ref{max_long_ft}, we obtain:
\begin{equation}
\label{flu_mom_long_ft_res}
\begin{aligned}
\left(1-\sum_sP_s\right)a_2 = \frac{1}{2\omega_2}\sum_s P_s PC_{s2}
\end{aligned}
\end{equation}

\subsubsection{\label{sec:Fourier_EMW}EM Waves in Fourier Space}

Writing equation \ref{flu_mom_trans_lin} in terms of Fourier modes, we obtain:
\begin{equation}
\begin{aligned}
  &\frac{1}{2}\sum_{+,-} \left( -i\omega_\pm\vec{v}_{s\pm} - i \mu_s^{-1}\omega_\pm \vec{a}_{\pm}
    - s_s\omega_{cs}\vec{v}_{s\pm}\times\hat{z} \right) e^{i\psi_\pm} + cc = \\
&-\frac{1}{4}[ik_0v_{s2}\vec{v}_{s0}e^{i\psi_+}+ik_0v_{s2}^*\vec{v}_{s0}e^{i\psi_-}\\
&+\mu_s^{-1}(ik_0v_{s2}\vec{a}_0e^{i\psi_+}+ik_0\vec{a}_0v_{s2}^*e^{i\psi_-})]+ cc\\
%&-\frac{1}{4}ik_0 \left( e^{i\psi_+}v_{s2}\vec{v}_{s0} - e^{i\psi_-}v_{s2}\vec{v}_{s0}^* \right)
%+\frac{1}{4}ik_0\mu_s^{-1} \left( e^{i\psi_+}v_{s2}\vec{a}_{0} - e^{i\psi_-}v_{s2}\vec{a}_{0}^* \right) + cc
\end{aligned}
\end{equation}
Let $Z_{y+}, Z_{y-}$ denote $Z_{y}$ and $Z_{y}^*$, respectively, where $Z$ denotes an amplitude, frequency or wavelength, and $y$ denotes a subscript containing the mode and plasma species (if applicable) of Z. This allows us to write generic equations for the + and - waves.  Selecting only resonant terms we obtain: 
\begin{equation}
\label{flu_mom_trans_ft_res}
\begin{aligned}
\omega_\pm(\vec{v}_{s\pm}+\mu_s^{-1}\vec{a}_{\pm})-is_s\omega_{cs}\vec{v}_{s\pm}\times\hat{z}=\frac{k_0}{2}v_{s2\pm}(\vec{v}_{s0} + \mu_s^{-1}\vec{a}_{0})
\end{aligned}
\end{equation}
Finally, equation \ref{max4_perp_lin}, once written in terms of Fourier modes, becomes:
\begin{equation}
\begin{aligned}
& \frac{1}{2}\sum_{+,-} \left( (-\omega_\pm^2+c^2k_\pm^2)\vec{a}_\pm - \sum_s\omega_{ps}^2\mu_s\vec{v}_\pm \right) e^{i\psi_\pm} + cc = \\
&\sum_s\omega_{ps}^2\frac{\mu_s}{4n_{seq}}(\vec{v}_{s0}n_{s2}e^{i\psi_+} + \vec{v}_{s0}n_{s2}^*e^{i\psi_-}) + cc
\end{aligned}
\end{equation}
Keeping terms resonant with $\psi_\pm$ and eliminating $n_{s2}$ gives
\begin{equation}
\label{max_trans_ft_res}
(-\omega_\pm^2+k_\pm^2c^2)\vec{a}_{\pm}-\sum_s\omega_{ps}^2\mu_s\vec{v}_{s\pm}=\frac{1}{2}\frac{k_{2\pm}}{\omega_{2\pm}}\sum_s\omega_{ps}^2\mu_sv_{s 2\pm}\vec{v}_{s 0}
\end{equation}
Using equation \ref{flu_mom_long_ft_vel} to eliminate $v_{s2}$ from equations \ref{flu_mom_trans_ft_res} and \ref{max_trans_ft_res}, keeping only terms up to second order we are left with the following equations, where we restate the plasma-wave equation for convenience:
\begin{equation}
\label{flu_mom_trans_ft_fnl}
\begin{aligned}
\vec{v}_{s\pm}+\mu_s^{-1}\vec{a}_{\pm}-i\beta_{s\pm}\vec{v}_{s\pm}\times\hat{z}=-K_{s\pm}a_{2\pm}(\vec{v}_{s0} + \mu_s^{-1}\vec{a}_{0})
\end{aligned}
\end{equation}
\begin{equation}
\label{max_trans_ft_fnl}
(-\omega_\pm^2+k_{\pm}^2c^2)\vec{a}_{\pm}-\sum_s\omega_{ps}^2\mu_s\vec{v}_{s\pm}=-\frac{k_{2\pm}\omega_{2\pm}}{2}\sum_sP_{s\pm}a_{2\pm}\vec{v}_{s 0}
\end{equation}
\begin{equation}
\begin{aligned}
\label{flu_mom_long_ft_res2}
(1-\sum_sP_s)a_2=\frac{1}{2\omega_2}\sum_sP_sPC_{s2}
\end{aligned}
\end{equation}
$K_{s\pm}=\frac{k_0\omega_{2\pm}^{2}P_{s\pm}}{2\mu_s\omega_\pm\omega_{ps}^2}$, $\beta_{s\pm}=s_s\frac{\omega_{cs}}{\omega_\pm}$ and $P_{s\pm}=\frac{\omega_{ps}^2}{\omega_{2\pm}^2-\gamma_sk_{2\pm}^2v_{Ts}^2}$. $\omega_{2+}=\omega_2$, $\omega_{2-}=\omega_2^*$, and similarly for $k_{2\pm}$. The equations for wave 0 are equivalent to those for the $\pm$ waves, neglecting second order terms.

At this point, the remaining task is to solve for $\vec{v}_{s\pm}$ in terms of $\vec{a}_{\pm}$, $a_2$, and wave 0 quantities. We will finally arrive at a $5\times 5$ system for $\vec{a}_+$, $\vec{a}_-^*$, and $a_2$, which includes both the linear waves and parametric coupling to wave 0. For magnetised waves, this is most easily done in a rotating coordinate system, where $R$ and $L$ circularly-polarised waves are the linear light waves.

\subsection{\label{sec:L_R_coords}Left-Right Co-ordinate System}

It is convenient when dealing with Fourier amplitudes to formulate vectors in terms of right and left polarised co-ordinates, which are defined in terms of cartesian co-ordinates as follows:
\begin{equation}
\begin{aligned}
&\hat{R}=\frac{1}{\sqrt{2}}(\hat{x}+i\hat{y})\\
&\hat{L}=\frac{1}{\sqrt{2}}(\hat{x}-i\hat{y})
\end{aligned}
\end{equation}
In condensed notation,
\begin{equation}
\hat{\sigma}=\frac{1}{\sqrt{2}}(\hat{x}+i\sigma \hat{y})
\end{equation}
where $\sigma=+1, -1$ for the right and left-polarised basis vectors, respectively. We define the dot product such that $\vec{a}\cdot\vec{b}=\sum_ia_ib_i^*$. Thus, dot products do not commute: $\vec{a}\cdot\vec{b}=(\vec{b}\cdot\vec{a})^*$. This normalisation ensures $\hat{\sigma}\cdot\hat{\sigma}= 1$.
Using this convention, any vector can be re-written in terms of right and left polarised unit vectors and amplitudes.
 Consider, for example, the physical velocity vector $\vec{v}_\perp$, where we explicitly indicate Fourier amplitudes with subscript $f$:
\begin{equation}
\begin{aligned}
\label{circ_coord_example}
\vec{v}_\perp &= (\hat{x}v_{fx}+\hat{y}v_{fy})e^{i\psi}+cc\\
&=\frac{1}{\sqrt{2}}((\hat{R}+\hat{L})v_{fx}+i(\hat{L}-\hat{R})v_{fy})e^{i\psi}+cc\\
&=\frac{1}{\sqrt{2}}(\hat{L}(v_{fx}+iv_{fy})+\hat{R}(v_{fx}-iv_{fy}))e^{i\psi}+cc\\
&=\frac{1}{\sqrt{2}}(v_{fL}\hat{L}+v_{fR}\hat{R})e^{i\psi}+cc\\
&=e^{i\psi} \sum_\sigma v_{f\sigma}\hat{\sigma}+cc
\end{aligned}
\end{equation}
Note that $\vec v \cdot\hat\sigma= 2^{-1/2}e^{i\psi}(v_{x}-i\sigma v_{y})=e^{i\psi}v_{f\sigma}+cc$.  As an explicit example, for an R wave with $v_{fR}=V$ real and $v_{fL}=0$, $\vec{v}_\perp = 2^{1/2}V(\cos\psi, -\sin\psi)$.  At fixed $z$, $\vec v_{\perp}$ rotates clockwise as time increases when looking toward $-\hat z$, which is opposite to $\vec B_{eq}$. We therefore follow the convention used by Stix \cite{stix}, in which circular polarization is defined relative to $\vec B_{eq}$ and not $\vec k$.

We use the result given in the last line of equation \ref{circ_coord_example} to produce the definition of a dot product of two vectors in Fourier space in this coordinate system. Consider the vectors $\vec{v}$ and $\vec{a}$:
\begin{equation}
\begin{aligned}
\vec{v}.\vec{a}=e^{i(\psi_i-\psi_j^*)}(v_{fRi}a_{fRj}^*+v_{fLi}a_{fLj}^*)+cc
\end{aligned}
\end{equation}
where the subscripts $i, j$ are the wave indices.

\subsection{\label{sec:L_R_coords_EMW}EM Waves in Left-Right Coordinates}

Taking $\hat\sigma\cdot$ (equations \ref{flu_mom_trans_ft_fnl} and \ref{max_trans_ft_fnl}), we obtain
\begin{equation}
\label{mom_trans_r_and_l}
\begin{aligned}
(1+\sigma\beta_{s\pm})v_{s \pm \sigma}+\mu_s^{-1}a_{\pm\sigma}=-K_{s\pm}\left(\mu_s^{-1}a_{0\sigma}+v_{s0\sigma}\right)a_{2\pm}
\end{aligned}
\end{equation}
\begin{equation}
\label{max_trans_r_and_l}
(\omega_\pm^2-k_\pm^2c^2)a_{\pm\sigma}+\sum_s\omega_{ps}^2\mu_s v_{s \pm\sigma}=\frac{k_{2\pm}\omega_{2\pm}}{2}\sum_s P_{s\pm} a_{2\pm} v_{s 0\sigma}
\end{equation}
respectively, where $a_{\pm\sigma}\equiv \vec{a}_\pm\cdot\hat{\sigma}$. The definitions of $v_{s\pm\sigma}$, $v_{s0\sigma}$ and $a_{0\sigma}$ are analogous to that of $a_{\pm\sigma}$. We now have uncoupled equations for $(a_{\pm\sigma}, v_{s\pm\sigma})$ which is the advantage of using rotating coordinates. This is unlike the original x and y coordinates, which are coupled due to the $\vec{v}\times\vec{B}$ force. For the pump wave, we have these equations with subscript $\pm \rightarrow 0$ and set the RHS to 0. Thus
\begin{equation}
\label{v_0pm_sub}
v_{s0\sigma} = -\frac{1}{\mu_s(1+\sigma\beta_{s0})} a_{0\sigma}
\end{equation}
Rearranging equation \ref{mom_trans_r_and_l} to obtain an expression for $v_{s\pm\sigma}$
 \begin{equation}
 \label{mom_trans_r_and_l_vel}
 (1+\sigma\beta_{s\pm})v_{s\pm\sigma} = -\mu_s^{-1}a_{\pm\sigma}-\frac{\sigma K_{s\pm}\beta_{s0}}{\mu_s(1+\sigma\beta_{s0})} a_{0\pm\sigma}a_{2\pm}
 \end{equation}
Substituting this into equation \ref{max_trans_r_and_l}, and moving parametric coupling terms to the right-hand side, we obtain:
\begin{equation}
\label{em_modes_disp}
D_{\pm\sigma}a_{\pm\sigma}=-\Delta_{\pm\sigma 2}a_{0\sigma}a_{2\pm}
\end{equation}
where
\begin{equation}
\begin{aligned}
D_{\pm\sigma} &= \omega_\pm^2-k_\pm^2c^2-\sum_s\frac{\omega_{ps}^2}{1+\sigma\beta_{s\pm}} \\
\Delta_{\pm\sigma 2} &= \frac{\omega_{2\pm}}{2}\sum_s\frac{P_{s\pm}}{\mu_s}\frac{1}{1+\sigma\beta_{s0}} \left(k_{2\pm}-k_0\frac{\omega_{2\pm}}{\omega_\pm}\frac{\sigma\beta_{s0}}{1+\sigma\beta_{s\pm}} \right)
\end{aligned}
\end{equation}
This has the desired form, where wave amplitudes are written only in terms of $a$'s, not $v$'s. For no B field, all $\beta$'s are zero, and the parametric coupling coefficient $\Delta_{\pm\sigma2} \propto k_{2\pm}$, the usual unmagnetised result. To explain the notation, $D_{+R}$ gives the linear dispersion relation for the scattered upshifted R wave, and $\Delta_{+R2}$ is the parametric coupling coefficient for that wave and wave 2 (the plasma wave). Please see the parametric dispersion relation Eq.\ \ref{mat_coup} below.

\subsection{\label{sec:L_R_coords_ESW}Plasma Waves in Left-Right Coordinates}

Writing the $PC_{s2}$ term in equation \ref{flu_mom_long_ft_res2} in terms of right and left circularly polarised waves, we obtain:
\begin{equation}
\begin{aligned}
&PC_{s2} = -k_-^*(v_{s0R}a_{-R}^*+v_{s0L}a_{-L}^*)+k_0(v_{s-R}^*a_{0R}+v_{s-L}^*a_{0L})\\
&+k_+(v_{s0R}^*a_{+R}+v_{s0L}^*a_{+L})-k_0(v_{s+R}a_{0R}^*+v_{s+L}a_{0L}^*)
\end{aligned}
\end{equation}
Substituting for $\vec{v}_{s0}$ using equation \ref{v_0pm_sub}, and $\vec{v}_{s\pm}$ using equation \ref{mom_trans_r_and_l_vel}
\begin{equation}
\begin{aligned}
 &-\mu_sPC_{s2} = a_{0R}a_{-R}^*\left(\frac{k_0}{1+\beta_{s-}^*} - \frac{k_-^*}{1+\beta_{s0}}\right)+\\
 &a_{0L}a_{-L}^*\left(\frac{k_0}{1-\beta_{s-}^*} - \frac{k_-^*}{1-\beta_{s0}}\right)\\
 &+ a_{0R}^*a_{+R}\left(-\frac{k_0}{1+\beta_{s+}} + \frac{k_+}{1+\beta_{s0}}\right)\\
 &+a_{0L}^*a_{+L}\left(-\frac{k_0}{1-\beta_{s+}} + \frac{k_+}{1-\beta_{s0}}\right)
\end{aligned}
\end{equation}
Equation \ref{flu_mom_long_ft_res2} can now be written in a more condensed form:
\begin{equation}
\label{es_mode_disp}
D_2 a_2=- \sum_\sigma  \left(\Delta_{2 + \sigma}a_{0\sigma}^*a_{+\sigma} +\Delta_{2 - \sigma}a_{0\sigma}a_{-\sigma}^*\right)
\end{equation}
\begin{equation}
D_2=1-\sum_s P_s
\end{equation}
\begin{equation}
\Delta_{2+\sigma}=\frac{1}{2\omega_2}\sum_s \frac{P_s}{\mu_s}\left(\frac{k_{+}}{1+\sigma\beta_{s0}}-\frac{k_{0}}{1+\sigma\beta_{s+}}\right)
\end{equation}
\begin{equation}
\Delta_{2-\sigma}=\frac{1}{2\omega_2}\sum_s \frac{P_s}{\mu_s}\left(-\frac{k_{-}^*}{1+\sigma\beta_{s0}}+\frac{k_{0}}{1+\sigma\beta_{s-}^*}\right)
\end{equation}
We now have a plasma-wave relation involving just $a$'s.

\subsection{\label{sec:Parametric_Disp_Rel}Parametric Dispersion Relation}

Equations \ref{em_modes_disp} (really 4 equations: equation \ref{em_modes_disp} and its complex conjugate for $\sigma=R,L$) and \ref{es_mode_disp} form a system of 5 linear equations, which can be summarised in matrix form:
\begin{equation}
\label{mat_coup}
\begin{aligned}
&\begin{bmatrix}
D_{+R}&0&0&0&\Delta_{+R2}a_{0R}\\
0&D_{-R}^*&0&0&\Delta_{-R2}^*a_{0R}^*\\
0&0&D_{+L}&0&\Delta_{+L2}a_{0L}\\
0&0&0&D_{-L}^*&\Delta_{-L2}^*a_{0L}^*\\
\Delta_{2+R}a_{0R}^*&\Delta_{2-R}a_{0R}&\Delta_{2+L}a_{0L}^*&\Delta_{2-L}a_{0L}&D_{2}\\
\end{bmatrix}
\begin{bmatrix}
a_{+R}\\
a_{-R^*}\\
a_{+L}\\
a_{-L^*}\\
a_{2}\\
\end{bmatrix}\\
&=0
\end{aligned}
\end{equation}
The structure of this matrix matches our physical understanding of plasma-wave dispersion relations: the diagonal terms are independent of $a$, and give rise to linear waves. The off-diagonal terms are all proportional to $a_0$ and represent parametric coupling between the two daughter waves, one e/m and the e/s plasma wave.  Nonzero solutions exist when the determinant is zero, which gives the parametric dispersion relation including the pump light wave in the equilibrium. This is analogous to Drake \cite{Drake74}, but generalized to include a background magnetic field, and specialized to our 1D geometry and fluid instead of kinetic plasma-wave response.

The parametric dispersion relation couples a pump and scattered e/m wave of the same R or L polarization. Consider the case where there is only one pump wave: i.e. either $a_{0R}=0$ or $a_{0L}=0$. Taking $a_{0R}=0$ for definiteness, waves $a_{-R}$ and $a_{-R}^*$ decouple from the dispersion relation, leaving the following dispersion matrix:
\begin{equation}
\label{mat_coup_simp}
\begin{bmatrix}
D_{+L}&0&\Delta_{+L2}a_{0L}\\
0&D_{-L}^*&\Delta_{-L2}^*a_{0L}^*\\
\Delta_{2+L}a_{0L}^*&\Delta_{2-L}a_{0L}&D_{2}\\
\end{bmatrix}
\begin{bmatrix}
a_{+L}\\
a_{-L}^*\\
a_{2}\\
\end{bmatrix}=0
\end{equation}
Setting the determinant to 0 gives
\begin{equation}
D_{+L}D_{-L}^*D_2=|a_{0L}|^2(D_{+L}\Delta_{2-L}\Delta_{-L2}^*+D_{-L}^*\Delta_{2+L}\Delta_{+L2})
\end{equation}
$a_{0L}=0$ then gives the three linear dispersion relations for the upshifted L, downshifted L, and plasma waves: $D_{+L}=0$, $D_{-L}=0$, or $D_2=0$.  $a_{0L} \neq 0$ couples the linear waves and gives parametric interaction.

\section{\label{sec:mag_free_waves}Impact of External B Field on Free Waves}

This section considers the linear or free waves, with $a_0=0$.  Let $a_1$ be either $a_+$ or $a_-$ in equation \ref{mat_coup} to obtain the free-wave dispersion relation:
\begin{equation}
\label{mat_no_coup}
\begin{bmatrix}
D_{1L}^*&0&0\\
0&D_{1R}^*&0\\
0&0&D_{2}\\
\end{bmatrix}
\begin{bmatrix}
a_{1L}^*\\
a_{1R}^*\\
a_{2}\\
\end{bmatrix}=0
\end{equation}
$\vec{a}\neq 0$ solutions exist if the determinant of this matrix equals 0. This gives rise to the following dispersion relations, for a single ion species. For the e/m waves, with $a_2=0$, we have $D_{1L}D_{1R}=0$, which gives
\begin{equation}
\label{EM_disp_free}
\omega_1^2 = k_1^2c^2 + \frac{\omega_{pe}^2}{1-\sigma\frac{\omega_{ce}}{\omega_1}} + \frac{\omega_{pi}^2}{1+\sigma\frac{\omega_{ci}}{\omega_1}}
\end{equation}
For e/s waves, with $a_{1L}=a_{1R}=0$, we have $D_2=0$ and 
\begin{equation}
\label{PW_disp_free}
\omega_{2}^2 = \frac{\omega_{pe}^2}{1-\gamma_{e}\frac{k_2^2v_{Te}^2}{\omega_2^2}}+\frac{\omega_{pi}^2}{1-\gamma_{i}\frac{k_2^2v_{Ti}^2}{\omega_2^2}}
\end{equation}
Note that the background $B$ field has no effect at all on the e/s waves, for our geometry of $\vec k || \vec B_{eq}$.

\subsection{\label{sec:unmagnetised_waves}Waves in an unmagnetised Plasma}

By setting $\omega_{ce}=0$, we recover the unmagnetised dispersion relation for electromagnetic waves from equation \ref{EM_disp_free}:
\begin{equation}
\label{EMW}
\omega_1^2=c^2 k_1^2+\omega_{pe}^2 + \omega_{pi}^2
\end{equation}
The ion contribution is usually negligible.  Equation \ref{PW_disp_free} gives the electrostatic waves, with the conventional approximations, like neglecting ions for electron plasma waves (EPWs), being highly accurate.  Namely, we find the EPW for $\gamma_e=3$:
 \begin{equation}
\label{EPW_warm}
\omega_2^2=\omega_{pe}^2+3v_{Te}^2k_2^2
\end{equation}
and the ion acoustic wave (IAW) for $\gamma_e=1,\gamma_i=3$:
\begin{equation}
\label{IAW}
\omega_2^2 = \frac{Z_iT_e}{m_i}\left(\frac{1}{1+(k_2\lambda_{De})^2}+\frac{3T_i}{Z_iT_e}\right)k_2^2
\end{equation}
with $\lambda_{De}\equiv v_{Te}/\omega_{pe}$.  We must retain finite $T_e$ for an IAW to exist.

\subsection{\label{sec:mag_waves}Waves with Magnetic Field}

The dispersion relation for free electromagnetic waves in a magnetised plasma is given in equation \ref{EM_disp_free}.  As is usual in LPI literature, we view this as giving $\omega$ as a function of real $k$.  This gives a 4th order polynomial for $\omega$ with four real solutions, each of which corresponds to an e/m wave:
\begin{equation} \label{EM_disp_4poly}
\begin{aligned}
 &\omega^4 - \sigma(\omega_{ce}-\omega_{ci})\omega^3 - (c^2k^2+\omega_{ce}\omega_{ci}+\omega_{pe}^2+\omega_{pi}^2)\omega^2\\
 &+ \sigma(\omega_{ce}-\omega_{ci})c^2k^2\omega + \omega_{ce}\omega_{ci}c^2k^2 = 0
\end{aligned}
\end{equation}
Note one can solve this trivially in closed form for $k$ given $\omega$. In the following analysis, but not in the numerical solutions, we assume $Z_im_e/m_i \ll 1$, so we can drop $\omega_{pi}^2$ and set $\omega_{ce}-\omega_{ci} \rightarrow \omega_{ce}$. In order of descending frequency, these waves are: the right and left-polarised light waves, the whistler wave and the ion cyclotron wave (ICW). In addition to these waves, two electrostatic waves are obtained by solving equation \ref{PW_disp_free}: the EPW and the IAW.

Let us consider the high-frequency e/m waves, the light and whistler waves, where ion motion can be neglected: $\omega_{ci} \rightarrow 0$. In this case, equation \ref{EM_disp_4poly} becomes (removing one $\omega=0$ root)
\begin{equation}
\label{EM_disp_no_ions}
  \omega^3 - \sigma\omega_{ce}\omega^2 - (c^2k^2+\omega_{pe}^2)\omega +\sigma\omega_{ce}c^2k^2 = 0
\end{equation}
We assume $\omega_{pe} \gg \omega_{ce}$, which is typical in the HED regime. For light waves, we consider equation \ref{EM_disp_no_ions} for $\omega \gg \omega_{ce}$.  For $k=0$, we find
\begin{equation}
\omega(k=0) \approx \omega_{pe} + \frac{\sigma}{2}\omega_{ce}
\end{equation}
For all $k$ we write $\omega$ as $\omega(B_{eq}=0) \equiv (c^2k^2+\omega_{pe}^2)^{1/2}$ plus a correction:
\begin{equation}
\omega \approx \omega(B_{eq}=0) + \frac{\sigma}{2}\frac{\omega_{pe}^2}{\omega(B_{eq}=0)^2}\omega_{ce}
\end{equation}

\textbf{Whistler wave:} We can also solve equation \ref{EM_disp_no_ions} for the whistler wave, which has $\omega \leq \omega_{ce}$. We call this full set of roots for $\omega$ the whistler, though some authors only use this term for the small $k$ domain and use ``electron cyclotron wave'' when $\omega$ is near $\omega_{ce}$. We derive expressions for this wave by considering two limits: first, for $k \rightarrow 0$ (but still large enough that we can neglect ion motion, discussed below), we obtain:
\begin{equation}
\omega \approx \sigma \frac{c^2k^2}{\omega^2_{pe}}\omega_{ce}
\end{equation}
We restrict interest to $\omega>0$ waves, which for the whistler requires the R wave ($\sigma=1$):
\begin{equation}
\label{em_whistler_approx_small_w}
\omega \approx \frac{c^2k^2}{\omega^2_{pe}}\omega_{ce} \qquad \sigma=1
\end{equation}
Secondly, for $ck \gg \omega_{pe}$, we obtain:
\begin{equation}
\label{em_whistler_approx_large_k_w_leq_wce}
\omega \approx \omega_{ce}\left( 1 - \frac{\omega^2_{pe}}{c^2k^2} \right) \qquad \sigma=1
\end{equation}
For $\omega$ near $\omega_{ce}$, the whistler group velocity $d\omega/dk$ approaches zero. Since this is the relevant wave propagation speed for three-wave interactions, such a localized whistler wavepacket would propagate very slowly.  This impacts how stimulated whistler scattering evolves, and how to practically realize the process in experiments or simulations.

%[width=0.5\textwidth,grid,tics=10]{pictures/baum}
% \put (20,85) {\huge$\displaystyle\gamma$}
\begin{figure}[ht!]
	\subcaptionbox{\label{EMW_EPW_B_5000_with_analytic}}{\begin{overpic}[width=8.0cm, trim={0.5cm 0.1cm 0.3cm 0.3cm}, clip]{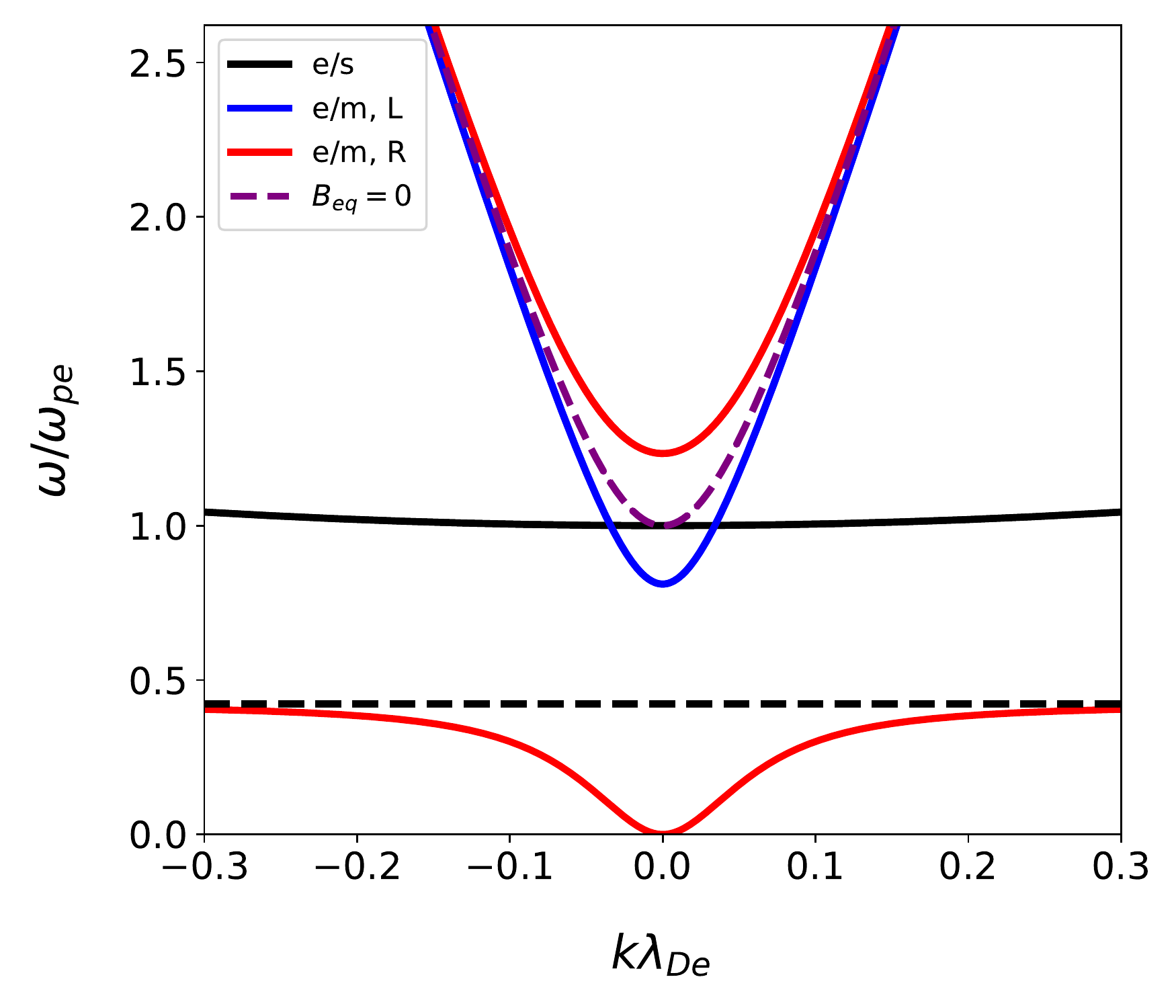}
	\put (20,30) {$\omega_{ce}$}
	\end{overpic}}
	\subcaptionbox{\label{IAW_ICW_B_5000_with_analytic}}{\includegraphics[width=8.0cm, trim={0.4cm 0.1cm 0.3cm 0.3cm}, clip]{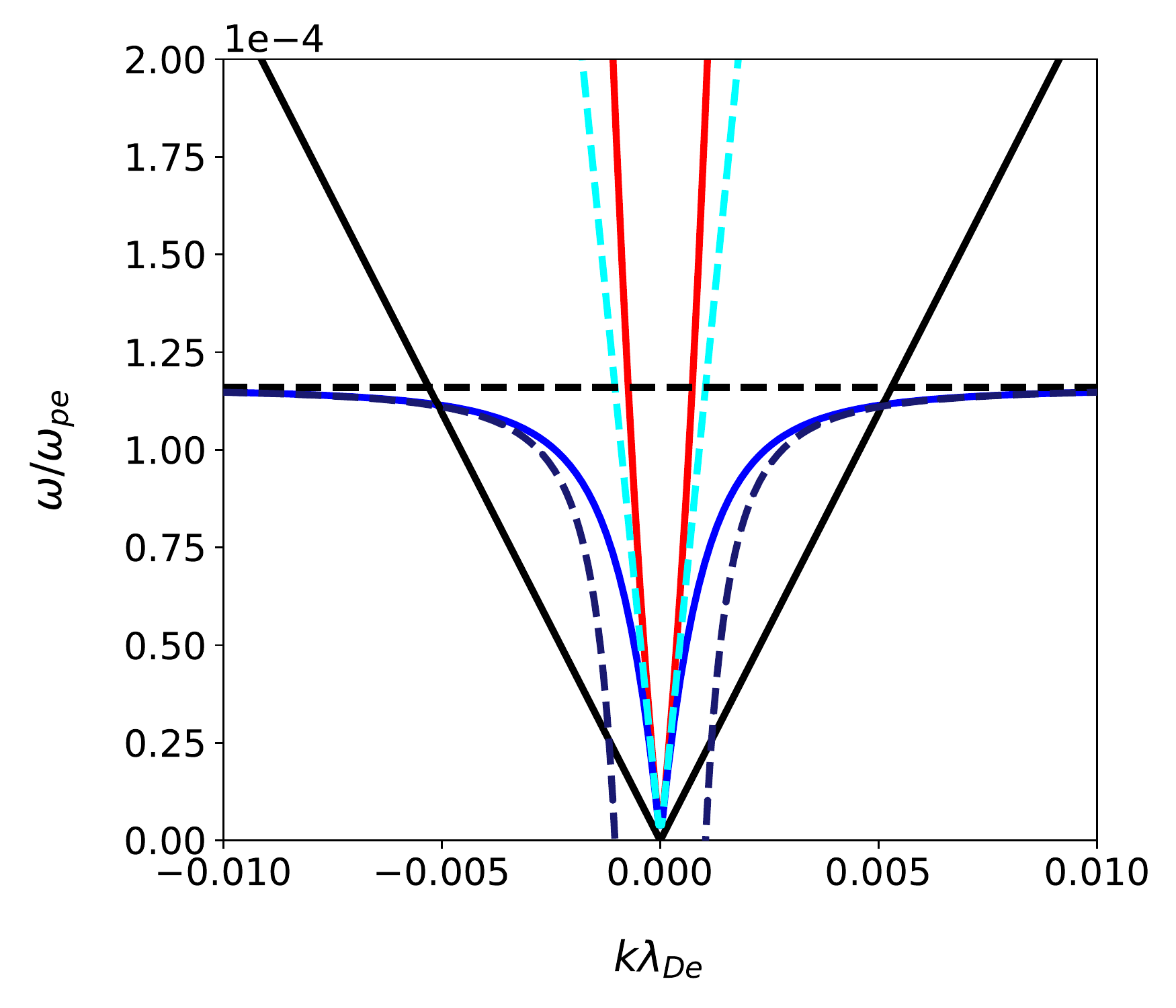}}
	\caption{Numerical solutions to the free-wave dispersion relations in a magnetized plasma, for the conditions in Table \ref{tablepara}. Red: right-polarised e/m, blue: left-polarised e/m, purple: unmagnetised e/m, and black: electrostatic.  Top: high-frequency waves, in decreasing order: e/m light, electron plasma, and whistler. The black dashed line lies at $\frac{\omega_{ce}}{\omega_{pe}}$. Bottom low-frequency waves: electrostatic ion acoustic wave, right-polarised whistler, and left-polarised ion cyclotron waves. Also plotted are the analytic approximations to the ion cyclotron wave for $ck \gg \omega_{pe}$ (dark blue) (equation \ref{ion_cyc_large_k}), which tends to $\frac{\omega_{ci}}{\omega_{pe}}$ (dashed black line), and $k\rightarrow0$, which yields the Alfven frequency (dashed cyan line), given in equation \ref{alfven}. }
%  \label{EMW_EPW_B_5000_with_analytic}

\end{figure}

The full numerical solutions of the dispersion relations for the whistler wave and the right and left-polarised light waves are shown in figure \ref{EMW_EPW_B_5000_with_analytic}. Note that here and throughout the rest of the paper, $\lambda_{De}$ is used to normalise $k$, as is customary for stimulated scattering. For large $k\lambda_{De}$, the whistler wave tends to $\omega=\omega_{ce}$, shown in figure \ref{EMW_EPW_B_5000_with_analytic} as a dashed black line.

\textbf{Ion cyclotron wave:} We now consider the ion cyclotron wave (ICW) which requires the retention of terms involving ion motion. As with the whistler wave, we consider two regimes. For $k\rightarrow 0$, we seek solutions with $\omega \propto k$, which gives
\begin{equation} \label{alfven}
\omega \approx v_Ak \qquad \sigma= -1 \textrm{ or } +1
\end{equation}
where the Alfven velocity, $v_A = c\frac{\omega_{ci}}{\omega_{pi}} = B/(\rho\mu_0)^{1/2}$. This solution applies for both values of $\sigma$, meaning there is both an R wave (the whistler, including ion motion), and an L wave (the ICW).  To see which is which, we need to take the opposite limit $ck \gg \omega_{pe}$, where we obtain two solutions with $\omega$ independent of $k$: $\omega=\omega_{ce}$ for $\sigma=1$ (the right-polarised whistler), and $\omega=\omega_{ci}$ for $\sigma=-1$ (the left-polarised ICW). Including the next correction term for the ICW gives
\begin{equation}
\label{ion_cyc_large_k}
\omega \approx \omega_{ci}\left(1-\frac{\omega^2_{pi}}{c^2k^2}\right) \qquad \sigma=1
\end{equation}
Figure \ref{EMW_EPW_B_5000_with_analytic} is re-plotted in figure \ref{IAW_ICW_B_5000_with_analytic} for $\omega \ll \omega_{pe}$ to show the IAW and ICW clearly. The ICW tends to $\omega = \omega_{ci}$, denoted by a dashed black line.
The numerical and approximate analytic solutions to the ICW dispersion relation are shown in figure \ref{IAW_ICW_B_5000_with_analytic} in blue and dark blue respectively. As can be seen from equation \ref{alfven}, at low $k\lambda_{De}$ the ICW approaches the Alfven frequency, which is represented by a dashed cyan line in figure \ref{IAW_ICW_B_5000_with_analytic}. For large values of $k\lambda_{De}$, the ICW frequency tends to $\omega_{ci}$, marked by a dashed black line.
The parameters used to plot the dispersion relations shown in figures (\ref{EMW_EPW_B_5000_with_analytic}-\ref{IAW_ICW_B_5000_with_analytic}) are given in table \ref{tablepara}. A plasma comprised of helium ions and electrons was considered.

\begin{table}[h!]
\begin{ruledtabular}
\begin{tabular}{|c|c|}
\centering
Quantity & Value \\
\hline
Z & 2\\
A&4\\
$T_e$&2 keV\\
$T_i$&1 keV\\
$\frac{\omega_{ce}}{\omega_{pe}}$ & 0.423
\end{tabular}
\end{ruledtabular}
\caption{\label{tablepara}Parameters used to plot dispersion relations.}
\end{table}

\begin{table}[!ht]
\centering
\begin{ruledtabular}
\begin{tabular}{|c|c|c|c|}
Laser wavelength [$\mu$m] & $n_e/n_{crit}$ & $n_e$ [$cm^{-3}$] & $B_{eq}$ [T]\\
\hline
0.351 (NIF) & 0.15 & $1.36\times 10^{21}$ & 5000\\
0.351 & 0.01 & $9.05\times 10^{19}$ &1290\\
10.6 (CO2) & 0.15 & $1.49 \times 10^{18}$  &166\\
10.6 & 0.01 & $9.92 \times 10^{16}$  &42.7\\
\end{tabular}
\end{ruledtabular}
\caption{\label{table_NIF_CO2_params}Electron densities and magnetic field strengths which correspond to the normalised parameters considered throughout this paper, for typical NIF and CO2 laser wavelengths.  $\frac{\omega_{ce}}{\omega_{pe}}=0.423$ in all cases.}
\end{table}

\subsection{\label{sec:faraday_rot}Faraday Rotation}

Three unique waves exist in an unmagnetised plasma, of which two are electrostatic (the electron plasma wave, (EPW) and the ion acoustic wave, (IAW)) and one is electromagnetic (light wave, with two degenerate polarisations). If the electromagnetic wave is linearly  polarised, it can be written as the sum of two circularly polarised waves of equal amplitude and opposite handedness (R and L waves). If an external B field, $\vec{B_{eq}}$ is applied, the R and L waves experience different indices of refraction and propagate with differing phase velocities. Consequently, the overall polarisation of the electromagnetic wave, found by summing the R and L waves, rotates as the electromagnetic wave propagates through the plasma. This is the well-known Faraday effect, which is briefly derived below.

An expression for the wavenumber of the electromagnetic wave can be obtained by rearranging equation \ref{EM_disp_free}.
\begin{equation}
\label{mag_free_modes_rearr_k1}
k_\sigma = \frac{\omega}{c} \left( 1-\frac{\omega_{pe}^2}{\omega^2(1-\sigma \frac{\omega_{ce}}{\omega})} \right)^\frac{1}{2}\
\end{equation}
Two first-order Taylor expansions of equation \ref{mag_free_modes_rearr_k1}, assuming $\omega\gg\omega_{ce}$, and $\omega\gg\omega_{pe}$ yield:
\begin{equation}
\label{k_approx}
k_\sigma \approx K - \sigma \Delta K
\end{equation}
where
\begin{equation}
K = \frac{\omega}{c}\left(1-\frac{\omega_{pe}^2}{2\omega^2}\right),
\qquad 
\Delta K = \frac{\omega_{pe}^2}{2\omega^2} \frac{\omega_{ce}}{c}
\end{equation}
Consider a linearly polarised plane electromagnetic wave. We can write the physical electric field $\vec E = Re[\vec E_F]$ as the sum of the electric fields of two circularly polarised waves with opposite handedness:
\begin{equation}
\label{lin_2_cir}
\vec{E}_F = \epsilon(\hat{R}e^{i\psi_R} + \hat{L}e^{i\psi_L}) \qquad \psi_{R,L} \equiv k_{R,L}z-\omega t
\end{equation}
Writing this in Cartesian co-ordinates,
\begin{equation}
\frac{2^{1/2}}{\epsilon}\vec{E}_F = \hat{x}(e^{i\psi_L}+e^{i\psi_R}) + i\hat{y}(e^{i\psi_L}-e^{i\psi_R})
\end{equation}
Assuming $\epsilon$ is real,
\begin{equation}
\begin{aligned}
 &\vec E = E(\cos\phi,-\sin\phi)\\
 &E = |2^{1/2}\epsilon \cos[(1/2)(k_L+k_R)z-\omega t]|\\
   &\phi = \frac{1}{2}(k_L-k_R)z = \Delta K z
 \end{aligned}
\end{equation}
At a fixed $z$, $\vec E$ always lies along the same line in the $xy$ plane, with its exact position varying in time.  As $z$ varies, the angle $\phi$ this line makes with respect to the $x$ axis increases at the rate
\begin{equation}
\frac{\partial \phi}{\partial z} = \Delta K = 16.8 \frac{n_e}{n_{crit}}B_{eq}\textrm{[T]  [deg/mm]}
\end{equation}
The final formula is in practical units.  We have introduced the critical density $n_{crit} \equiv (\epsilon_0m_e/e^2)\omega^2$, which is the usual definition for unmagnetised plasma.  When discussing LPI, $n_{crit}$ is for the pump wave $\omega_0$. Significant Faraday rotation is thus possible in current ICF platforms with modest B fields.  For instance, with $n_e/n_{crit}=0.1$ and $B_{eq}=10$ T, we obtain $\partial_z\phi = 16.8^\circ/$mm.  This could be used to diagnose $n_e$ (a common technique when feasible), and could affect LPI processes such as crossed-beam energy transfer.~\citep{Randall_1981,Kruer96,MichaelPRL_2009}

\section{\label{sec:M}Impact of external B field on Parametric Coupling}

We apply the above theory to magnetized LPI in HED relevant conditions, all for $\vec{k}||\vec{B_{eq}}||\hat{z}$.  We consider how the imposed field modifies stimulated Raman (SRS) and Brillouin (SBS) scattering, as well as stimulated whistler scattering (SWS) which only occurs in a background field. Recall $\vec k_i = k_i\hat z$ and we choose $k_{0}>0$. $k_{1}$ and $k_{2}$ can have either sign.  Let $c_i=$sign($k_{i}$) for $i=1,2$.  For all three parametric processes we discuss, ``forward scatter'' refers to the case where the scattered e/m wave propagates in the same direction as the pump $(c_1=+1)$, and ``backward scatter'' to the opposite case $(c_1=-1)$.  To satisfy $k$ matching, we cannot have both $c_1=-1$ and $c_2=-1$.  For SRS and SBS, $c_2$ must equal +1, but for SWS $c_2=-1$ is possible.

We do not consider growth rates, but focus instead on the ``kinematics'' of three-wave interactions, through the phase-matching conditions among free waves. We study the scattered e/m wave frequency $\omega_1$, since this is what escapes the plasma and is measured experimentally. As discussed in section \ref{sec:faraday_rot}, $\vec{B_{eq}}$ causes the R and L waves to propagate with different phase velocities. Therefore, a laser or other external source that imposes a linearly-polarised light wave of frequency $\omega_0$ couples to an R and an L wave in a magnetised plasma. For stimulated scattering, we are mostly interested in down-shifted scattered waves for which $\omega_1<\omega_0$, which have the same polarisation as the pump: an R or L pump couples to a down-shifted R or L scattered wave, respectively, hence $\sigma_1=\sigma_0$ which we sometimes denote as $\sigma$. We discuss SRS and SBS, which can be driven by either an R or L pump, and SWS, which can only be driven by an R pump (since the whistler wave is an R wave).  Table \ref{table_3wavesum} summarizes the processes we study.

\begin{table*}
\begin{ruledtabular}
  \begin{tabular}{|c|c|c|c|c|c|c|}
    Process & pump e/m wave & scattered e/m wave & plasma wave & geometries $(c_1,c_2)$ & $\omega_1$ range             & $n_e/n_{crit}$ range \\
    \hline
    SRS     & R,L           & R,L                & EPW         & (1,1), (-1, 1)        & $>\omega_{pe}$                & $<1/4$ \\
    SBS     & R,L           & R,L                & IAW         & (1,1), (-1, 1)        & $\gtrsim \omega_0-\omega_{pi}$ & $<1$ \\
    SWS     & R             & R-whistler         & EPW         & (1,-1), (-1, 1) & $<\omega_{ce}$               & $\gtrsim(1-\omega_{ce}/\omega_0)^2$ for $T_e=0$ \\
  \end{tabular}
  \end{ruledtabular}
  \caption{Summary of parametric processes we study.  L, R refer to left, right polarised e/m waves.}
  \label{table_3wavesum}
\end{table*}

\begin{figure}[ht!]
\centering
{\begin{overpic}[width=8.5cm, trim={0.4cm 0.2cm 0.35cm 0.35cm}, clip]{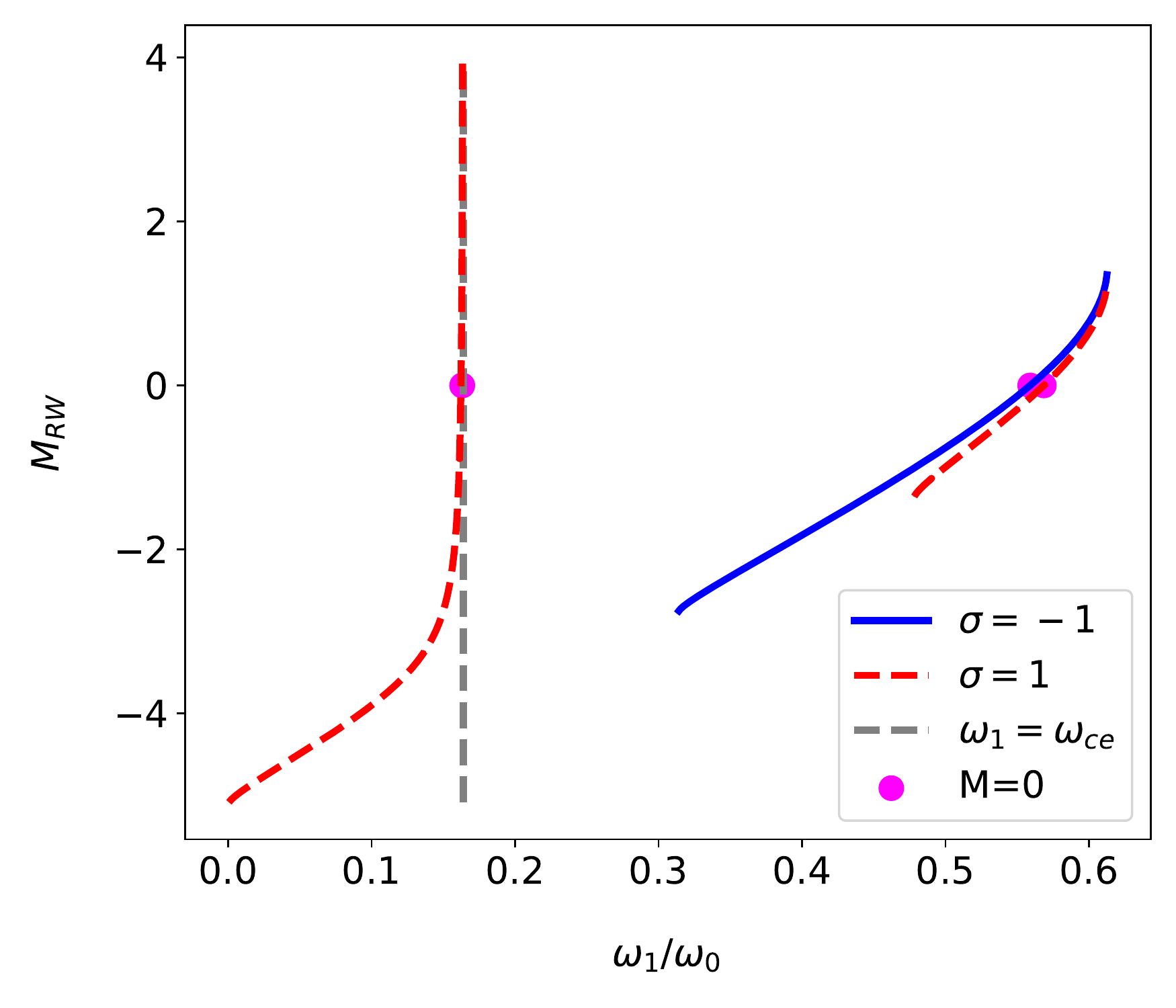}
\put (80,55) {SRS}
\put (25,50) {SWS}
\end{overpic}}
\caption{\label{MRW_back_scatter_R&L_pol}The dispersion relation for SRS and SWS, $M_{RW}$ is plotted vs.\ $\omega_1/\omega_0$. Its roots $M_{RW}=0$ are indicted by magenta points. This is for backscatter ($c_1=-1, c_2=1$) and the parameters of Table 1 plus $n_e/n_{crit}=0.15$. SWS is possible for a right polarised pump (red), but cannot occur when the pump is left polarised (blue).}
\end{figure}

\begin{figure}[ht!]
\centering
\includegraphics[width=8.5cm, trim={0.4cm 0.45cm 0.35cm 0.3cm}, clip]{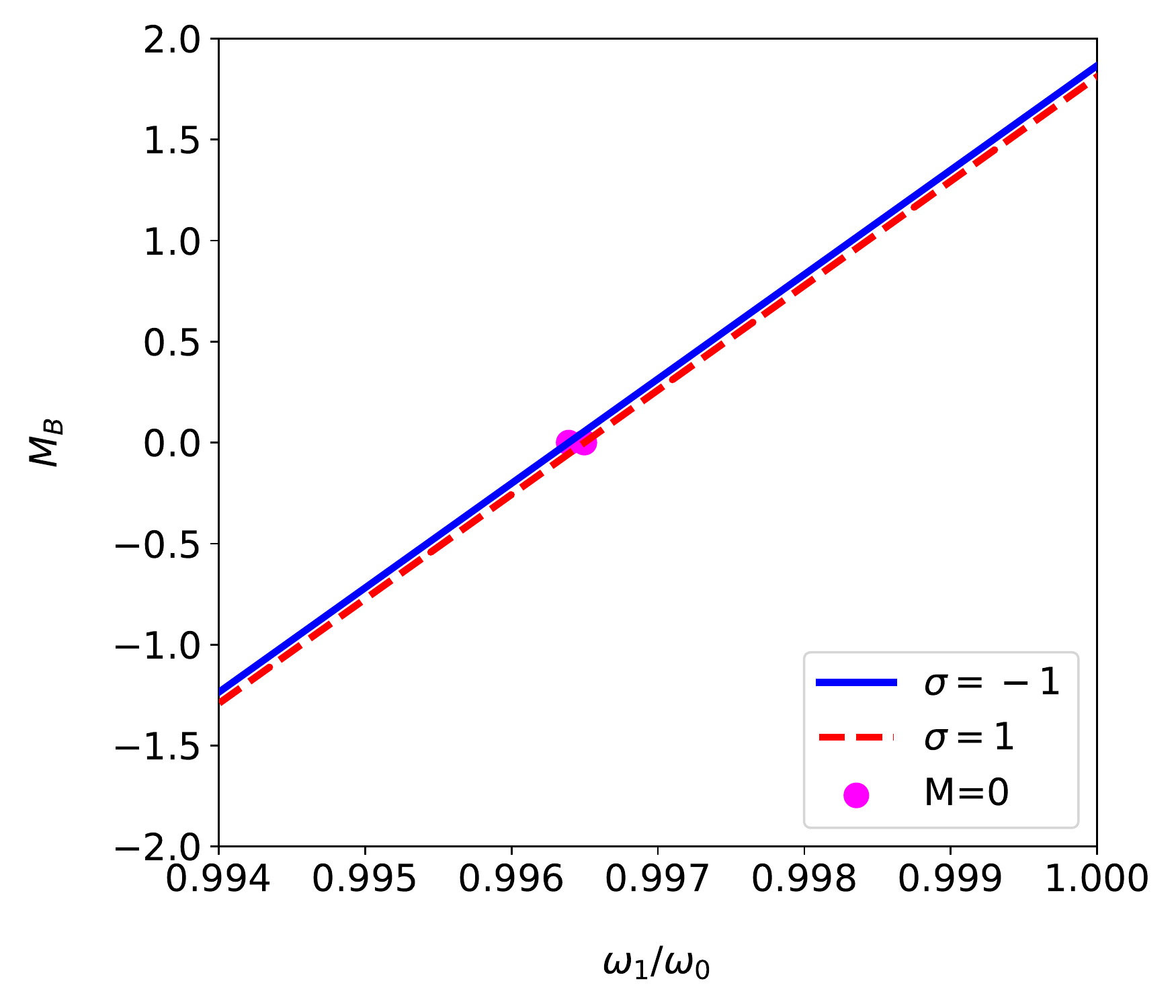}
\caption{The dispersion relation for SBS, $M_{B}$ is plotted vs.\ $\omega_1/\omega_0$, for the same parameters as Fig.\ \ref{MRW_back_scatter_R&L_pol}. Its roots $M_B=0$ are indicted by magenta points. The roots of $M_B$ occur at similar, but not identical $\omega_1/\omega_0$ for a left and right polarised pump.\label{MB_back_scatter_R&L_pol}}
\end{figure}

In order to derive a dispersion relation for $\omega_1$ in terms of known inputs, we begin with the identity $k_2=k_2$. We use $k$ matching to write $k_2=k_0-k_1$ on the left side, and the plasma-wave dispersion relation of interest to re-write the right side in terms of $\omega_2$. We then use the e/m dispersion relation to write $k_1$ in terms of $\omega_1$, and use $\omega$ matching to write $\omega_2=\omega_0-\omega_1$. 
For SRS and SWS this yields $k_0-k_1 = (\omega_2^2-\omega_{pe}^2)^{1/2}/v_{Te}3^{1/2}$. The same method is applied for SBS, where $k_2$ is written in terms of $\omega_2$ using the simple IAW dispersion relation, $\omega_2=c_s|k_2|$, for an approximate analysis (the numerical roots use the full e/s dispersion relation). That is, $c_s^2 = (Z_iT_e/m_i)(1 + 3T_i/Z_iT_e)$. The resulting dispersion relations can be summarised as follows:
\begin{equation}
\begin{aligned}
\label{M_general}
  &M_Y \equiv (1-\Omega_{pe}^2(1-\sigma_0\Omega_{ce})^{-1})^{1/2} \\
  &- c_1\Omega_1(1-\Omega_1^{-2}\Omega_{pe}^2(1-\sigma_1\Omega_{ce}/\Omega_1)^{-1})^{1/2} -P_Y=0
  \end{aligned}
\end{equation}
where $Y$ is either RW, for SRS and SWS, or B, for SBS.
For SRS and SWS $P_Y=P_{RW}=c_2 V_e^{-1}((1-\Omega_1)^2-\Omega_{pe}^2)^{1/2}$, where $V_e \equiv v_{Te}3^{1/2}/c$. For SBS, $P_Y=P_{B}=V_s^{-1}(1-\Omega_1)$, with $V_s \equiv c_s/c$. This is usually very small, with $10^{-3}$ a typical magnitude. $\Omega_X \equiv \omega_X/\omega_0$, where $X$ denotes any angular frequency subscript in equation \ref{M_general}.
The frequency of scattered light which satisfies phase matching is given by the roots of equation \ref{M_general}, which can be found by plotting $M_Y$ vs.\ $\Omega_1$. This is illustrated for SRS and SWS in figure \ref{MRW_back_scatter_R&L_pol}, and for SBS in figure \ref{MB_back_scatter_R&L_pol}, for the parameters given in table \ref{tablepara} and $n_e/n_{crit}=0.15$.

\begin{figure}[ht!]
\centering
\caption{The dispersion relations for SWS and SRS $(M_{RW})$ and SBS $(M_B)$ vs.\ electron density and scattered light frequency. $T_e=4$ keV, $T_i=2$ keV, $\omega_{ce}/\omega_{pe}=0.423$, and we consider backscatter $(c_1=-1, c_2=1)$. $M_{RW}$ is distinguished by cross-hatching. The roots of $M$ are plotted as black contours which have been labelled appropriately. Three other curves have been plotted: $n_e/n_{crit}=0.25$, the maximum density at which SRS occurs, $\omega_1=\omega_{ce}$, the maximum SWS frequency, and $n_e/n_{crit}\geq (1-\omega_{ce}/\omega_0)^2$, the \textit{minimum} density at which SWS can occur in a cold plasma. Note that $M_{RW}$ adheres to only the first two of these approximate analytic limits.\label{sws_contur_c1=-1_c2=1}}{\begin{overpic}[width=8.5cm, trim={0.4cm 0.4cm 0.4cm 0.3cm}, clip]{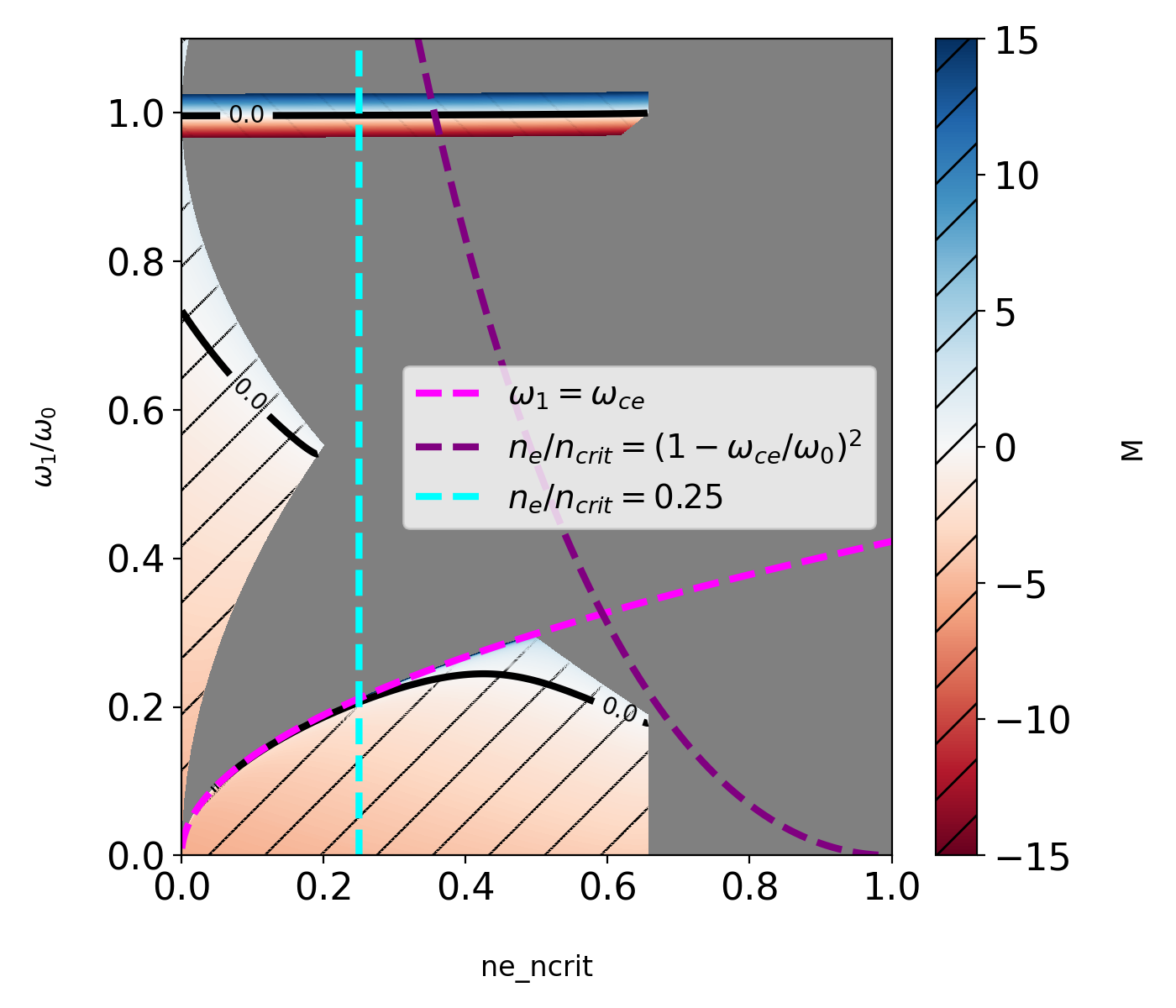}
\put (18,60) {SRS}
\put (32,22) {SWS}
\put (20,80) {SBS}
\end{overpic}}
\end{figure}

\begin{figure}[ht!]
\centering
\caption{As figure \ref{sws_contur_c1=-1_c2=1}, but for forward scatter $(c_1=c_2=1)$. Only SRS can occur for this geometry.  While SBS is kinematically possible, the ion wave has $k_2, \omega_2=0$, and SBS has 0 growth rate. Thus, the solution plotted is spurious. For this geometry, SWS is kinematically disallowed.\label{sws_contur_c1=1_c2=1}}{\begin{overpic}[width=8.5cm, trim={0.4cm 0.4cm 0.4cm 0.3cm}, clip]{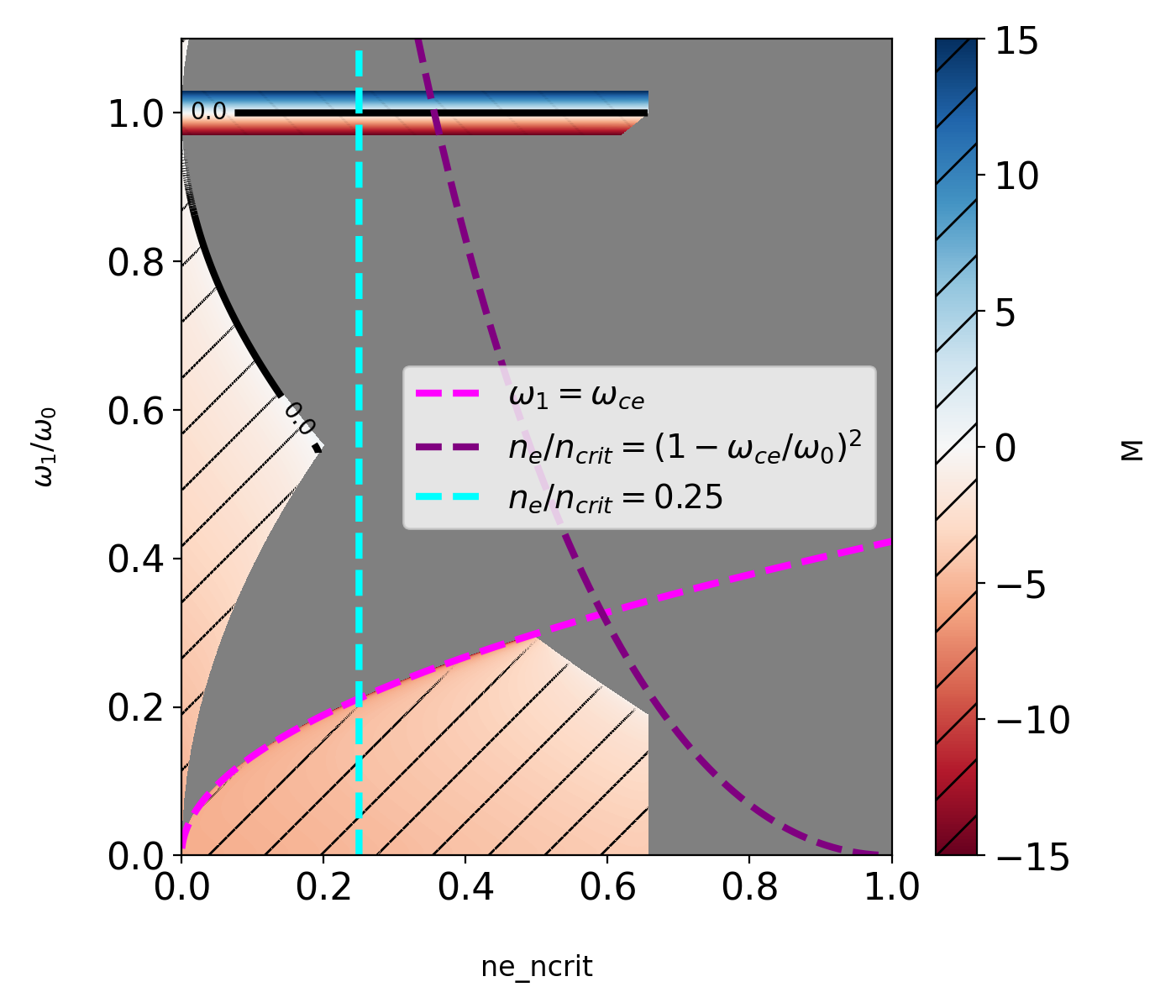}
\put (18,64) {SRS}
\put (20,80.25) {SBS}
\end{overpic}}
\end{figure}

\begin{figure}[ht!]
\centering
\caption{As figure \ref{sws_contur_c1=-1_c2=1}, but for $c_1=1$ and $c_2=-1$. For this geometry, phase matching is only satisfied for SWS, and unphysical SBS as in figure \ref{sws_contur_c1=1_c2=1}. As in figure \ref{sws_contur_c1=-1_c2=1}, $M_{RW}=0$ is only satisfied for densities above the minimum normalised electron density in a cold plasma,  $n_e/n_{crit}\geq (1-\omega_{ce}/\omega_0)^2$, which is plotted in purple.\label{sws_contur_c1=1_c2=-1}}{\begin{overpic}[width=8.5cm, trim={0.4cm 0.4cm 0.4cm 0.3cm}, clip]{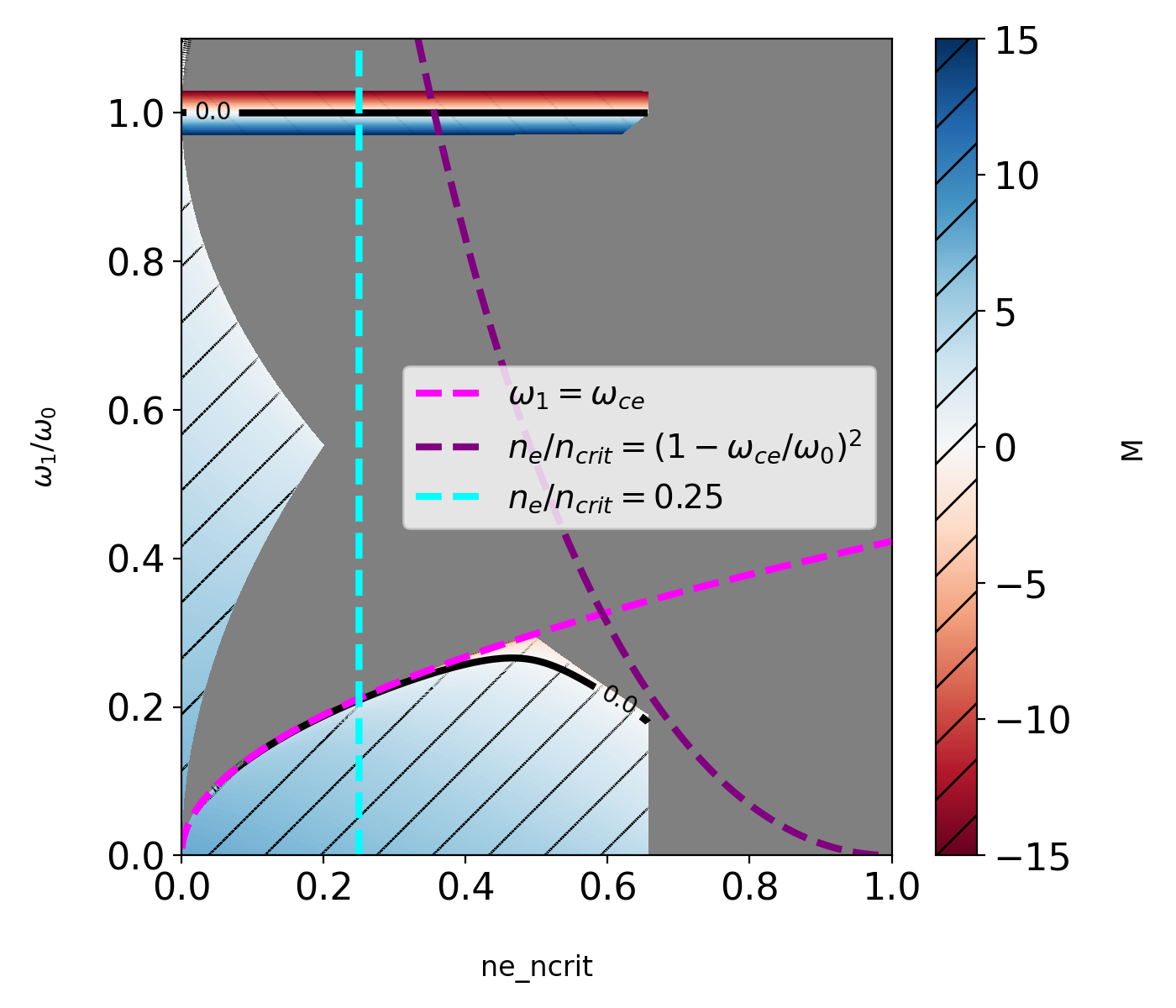}
\put (41,23) {SWS}
\put (20,80.25) {SBS}
\end{overpic}}
\end{figure}

The dispersion relations given in equation \ref{M_general} are plotted as a function of $\omega_1/\omega_0$ and $n_e/n_{crit}$ for scattering geometries $(c_1, c_2)=(-1, 1), (1, 1), (1, -1)$, in figures \ref{sws_contur_c1=-1_c2=1}, \ref{sws_contur_c1=1_c2=1} and \ref{sws_contur_c1=1_c2=-1}, respectively. 
The two dispersion relations, $M_{RW}$ and $M_{B}$ have been overplotted. To distinguish between them, $M_{RW}$ has been cross-hatched, whilst $M_{B}$ has not. The colour scale for M applies to both $M_{RW}$ and $M_{B}$. The regions of figures \ref{sws_contur_c1=-1_c2=1}, \ref{sws_contur_c1=1_c2=1} and \ref{sws_contur_c1=1_c2=-1} where M is not real are coloured gray. The regions of the plot where $M_{RW, B}\neq0$ serve only to illustrate the root-finding method employed: to ensure we have correctly identified roots, we check that $M_{RW, B}$ has changed sign.
The roots of M have been computed numerically and are plotted as black contours. These contours indicate whether SRS, SBS or SWS can occur for the geometry and plasma conditions considered, and illustrate the relationship between the normalised plasma density and scattered EMW frequency for each of these processes. The contours which correspond to a given parametric process are appropriately labelled. 

In figures \ref{sws_contur_c1=-1_c2=1} and \ref{sws_contur_c1=1_c2=1} a sharp decrease can be seen in the frequency of SRS scattered light with increasing plasma density. Also in figures \ref{sws_contur_c1=-1_c2=1} and \ref{sws_contur_c1=1_c2=-1}, the frequency of SWS scattered light rises with electron density before reaching a maximum, and falling.
It is often useful to obtain limits in parameter space beyond which phase matching cannot occur. For example, in an unmagnetised plasma, SRS is only possible for  $n_e/n_{crit}<0.25$. The region of parameter space in which SWS can occur is also restricted, as $\omega_1\leq\omega_{ce}$. Using the same method as for SRS, the following inequality is obtained for the normalised electron densities at which SWS can occur in a cold plasma: 
\begin{equation}
\label{sws_ncrit}
\frac{n_e}{n_{crit}}\geq (1-\omega_{ce}/\omega_0)^2
\end{equation}
These three limits are shown in figures \ref{sws_contur_c1=-1_c2=1}, \ref{sws_contur_c1=1_c2=1} and \ref{sws_contur_c1=1_c2=-1} in cyan, magenta and purple, respectively. Note that the contours for SRS and SWS always lie within $n_e/n_{crit}<0.25$ and $\omega_1\leq\omega_{ce}$ respectively, as expected.  SWS does \textit{not} respect Eq.\ \ref{sws_ncrit}, as discussed further below.
%\begin{figure}[ht!]
%\centering
%\begin{floatrow}
%\ffigbox[\FBwidth]{\caption{The dispersion relations for SWS and SRS $M_{RW}$, and that of SBS, $M_B$ are plotted as a function of electron density and scattered light frequency, normalised to the critical density and frequency, respectively. The electron and ion temperature are 4keV and 2keV, respectively, and $\omega_{ce}/\omega_{pe}=0.423$. $M_{RW}$ and $M_B$ are distinguished by cross-hatching which runs from bottom left to top right, and bottom right to top left, respectively. The roots of M are plotted as black contours which have been labelled appropriately. We only consider parametric processes with $\omega_1<\omega_0$, as only these processes can have a positive growth rate. For the geometry considered here, where $c_1=-1$ and $c_2=1$, phase matching is satisfied for SRS, SWS and SBS. Three trends have been plotted: $n_e/n_{crit}=0.25$, the maximum density at which SRS can occur in a cold plasma, $\omega_{ce}/\omega_0$, the maximum normalised frequency of SWS-scattered light, and  $n_e/n_{crit}\geq (1-\omega_{ce}/\omega_0)^2$, the minimum density at which SWS can occur in a cold plasma. Note that $M_{RW}$ adheres to the first two of these approximate analytic limits.}\label{sws_contur_c1=-1_c2=1}}{\begin{overpic}[width=8.5cm, trim={0.4cm 0.4cm 0.4cm 0.3cm}, clip]{srs_w_sbs_w_sws_phase_space_warm_wce_wpe=0pt423_c1=-1_c2=1_reduced_M_range.png}
%\put (18,60) {SRS}
%\put (32,22) {SWS}
%\put (20,80) {SBS}
%\end{overpic}}
%\end{floatrow}
%\end{figure}

\subsection{\label{sec:SRS}Stimulated Raman Scattering: SRS}

The dispersion relation for SRS is given by equation \ref{M_general}, where $c_2=1$. For a cold plasma with $V_e=0$, we find $\Omega_2 = \Omega_p$ always, so $\Omega_1 = 1-\Omega_p$.  This is true with or without a background field $B_{eq}$.  Thus, any effect of $B_{eq}$ on $\Omega_1$ is ``doubly small'', in that it also relies on thermal effects.  For no background field $\Omega_{ce}=0$, we obtain the usual solutions, which for $V_e\ll 1$ and $\Omega_p\ll 1$ are $\Omega_1 \approx 1-\Omega_p - (\Omega_p/2)V_e^2$ for $c_1=1$ (forward scatter), and $\Omega_1 \approx 1 - \Omega_p - (2/\Omega_p)V_e^2$ for $c_1=-1$ (backscatter).

\begin{figure*}[ht!]
  \caption{Phase-matching parallelograms for forward-SRS light for plasma conditions given in table \ref{tablepara}, with $n_e/n_{crit}=0.15$. The right and left-polarised e/m waves are plotted in red and blue, respectively, while the unmagnetised e/m wave and the electrostatic EPW are shown in purple and black, respectively. The phase-matching parallelograms are colour-coded according to the polarisation of the pump wave. The pump frequency $\omega_0$ is fixed in all cases, which gives slightly different $k_0$'s from the relevant dispersion relations. The scattered e/m frequencies $\omega_1$ are \textit{nearly} but not exactly the same, though this is very hard to see visually. The pump and scattered e/m waves have the same handedness.\label{SRS_f_ne_0.15_phase_match}}
  {\begin{overpic}[width=17.5cm, trim={0.0cm 0.0cm 0.0cm 0.0cm}, clip]{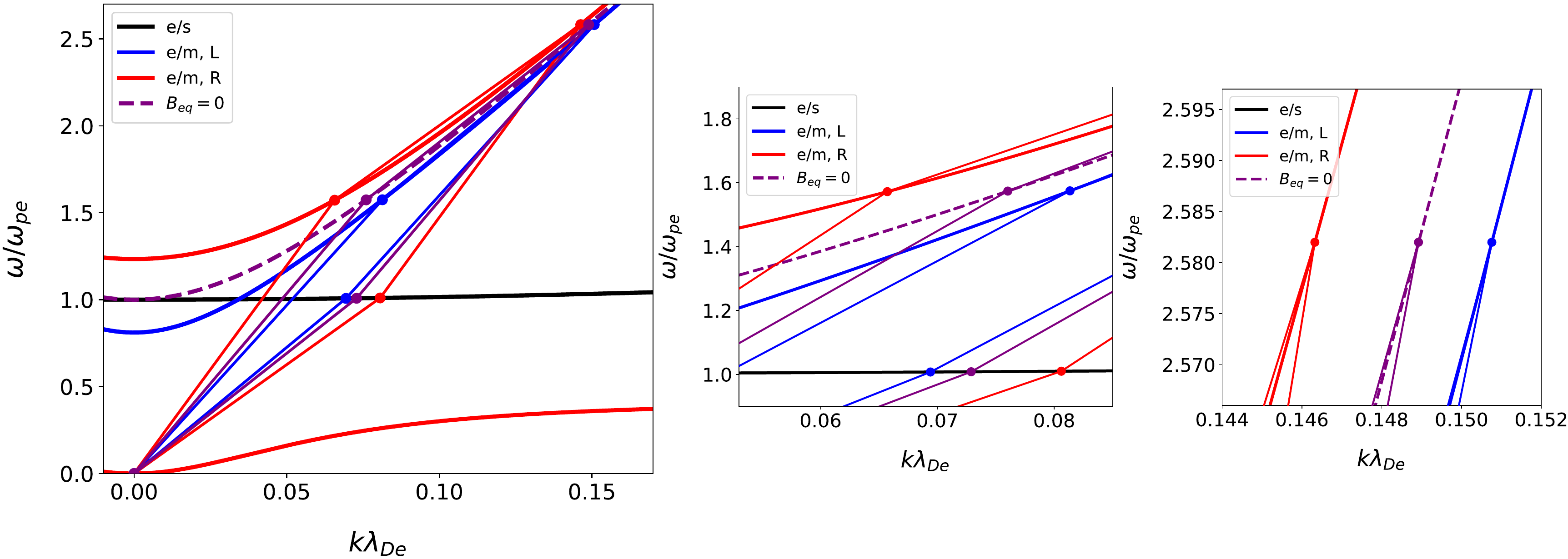}
\put (85.5,22) {Pump}
\put (56,27) {Scattered}
\put (58,15) {EPW}
\end{overpic}}
\end{figure*}

\begin{figure*}[ht!]
  \caption{Phase-matching parallelograms for backward-SRS light: otherwise same as Fig.\ \ref{SRS_f_ne_0.15_phase_match}.  \label{SRS_b_ne_0.15_phase_match}}
   {\begin{overpic}[width=17.5cm, trim={0.1cm 0.0cm 0.0cm 0.0cm}, clip]{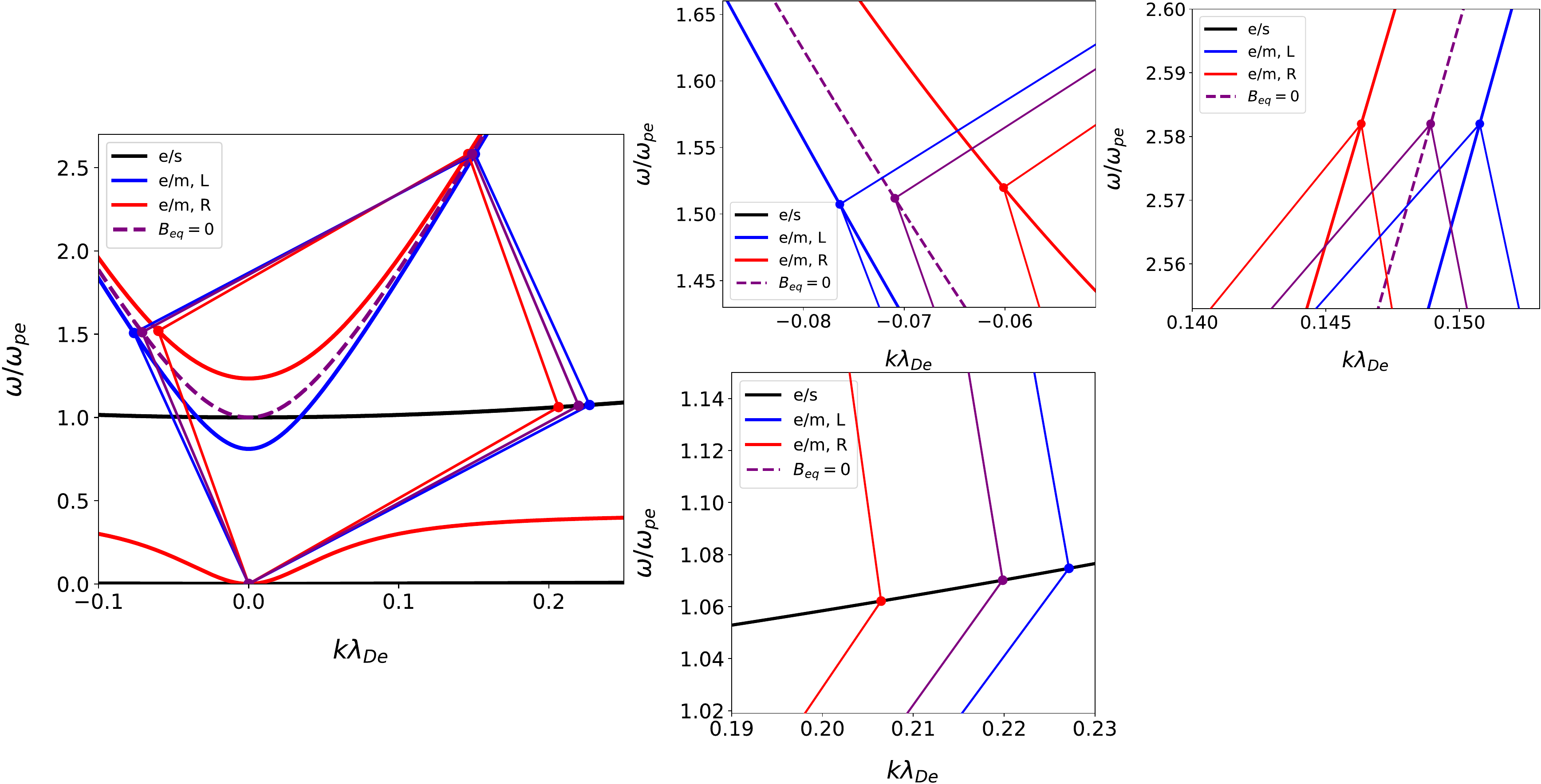}
  \put (80,41) {Pump}
\put (60,48) {Scattered}
\put (52,14) {EPW}
\end{overpic}}
\end{figure*}

Including a weak background field, we write $\Omega_1 \approx \Omega_{1U} + \delta\Omega_1$ where $\Omega_{1U}$ is the solution for $\Omega_{ce}=0$: $M[\Omega_{1U}, \Omega_{ce}=0]=0$. We have $M[\Omega_{1U}+\delta\Omega_1,\Omega_{ce}] \approx M[\Omega_{1U},0] + \delta\Omega_1(\partial M/\partial \Omega_1) + \Omega_{ce}\partial M/\partial\Omega_{ce}=0$, which gives $\delta\Omega_1 \approx \alpha\Omega_{ce}$ with $\alpha = -(\partial M/\partial \Omega_{ce})/(\partial M/\partial \Omega_1)$.  All partials are evaluated at $\Omega_1=\Omega_{1U}$ and $\Omega_{ce}=0$. One can find a formula for $\alpha$, but it is unilluminating.  We quote the result in the limit that $V_e\ll 1$ and $\Omega_p\ll 1$:
\begin{equation}
  \alpha \approx c_1 \left(2/\Omega_p^2+1/\Omega_p+2\right)^{\frac{1-c_1}{2}} \sigma_0 V_e^2\Omega_p^3
\end{equation}
The full numerical solution of $M_{RW}$ (see equation \ref{M_general}) is plotted in figures \ref{SRS_f_ne_0.15_phase_match} and \ref{SRS_b_ne_0.15_phase_match} for the plasma conditions given in table \ref{tablepara} and the first row of table \ref{table_NIF_CO2_params}. The frequencies, wave vectors and, if applicable, the polarisations of the e/m and e/s waves at which phase-matching conditions are met are illustrated by parallelograms. Specifically, figures \ref{SRS_f_ne_0.15_phase_match} and \ref{SRS_b_ne_0.15_phase_match} correspond to forward and back-SRS, respectively.

%\begin{figure}[ht!]
%  \caption{Phase-matching parallelograms for forward-SRS light: same as Fig.\ \ref{SRS_f_ne_0.15_phase_match} except $n_e/n_{crit,0}=0.05$.  [[DJS what is the benfit of this fig?  What does it show that the prior one doesn't?  Do we need it?  Or do we need all 3 panels? ]] \label{SRS_f_ne_0.15_phase_match}}
%  \includegraphics[width=18.0cm, trim={0.1cm 0.0cm 0.0cm 0.0cm}, clip]{SRS_forward_scatter_R_L_and_0_ne_0point05.pdf}
%\end{figure}

%\begin{figure}[ht!]
%  \caption{Phase-matching parallelograms for backward-SRS light: otherwise same as Fig.\ \ref{SRS_f_ne_0.15_phase_match} except $n_e/n_{crit,0}=0.05$. [[DJS do we need this fig?  What does it add vs. prior one?  And as before, if we need it do we need all 4 panels?]] \label{SRS_b_ne_0.15_phase_match}}
%  \includegraphics[width=18.0cm, trim={0.1cm 0.0cm 0.0cm 0.0cm}, clip]{SRS_backward_scatter_R_L_and_0_ne_0point05.pdf}
%\end{figure}

\begin{figure}[ht!]
\subcaptionbox{\label{SRS_f_vs_wce}}{\includegraphics[width=8.5cm, trim={0.1cm 0.5cm 0.3cm 0.3cm}, clip]{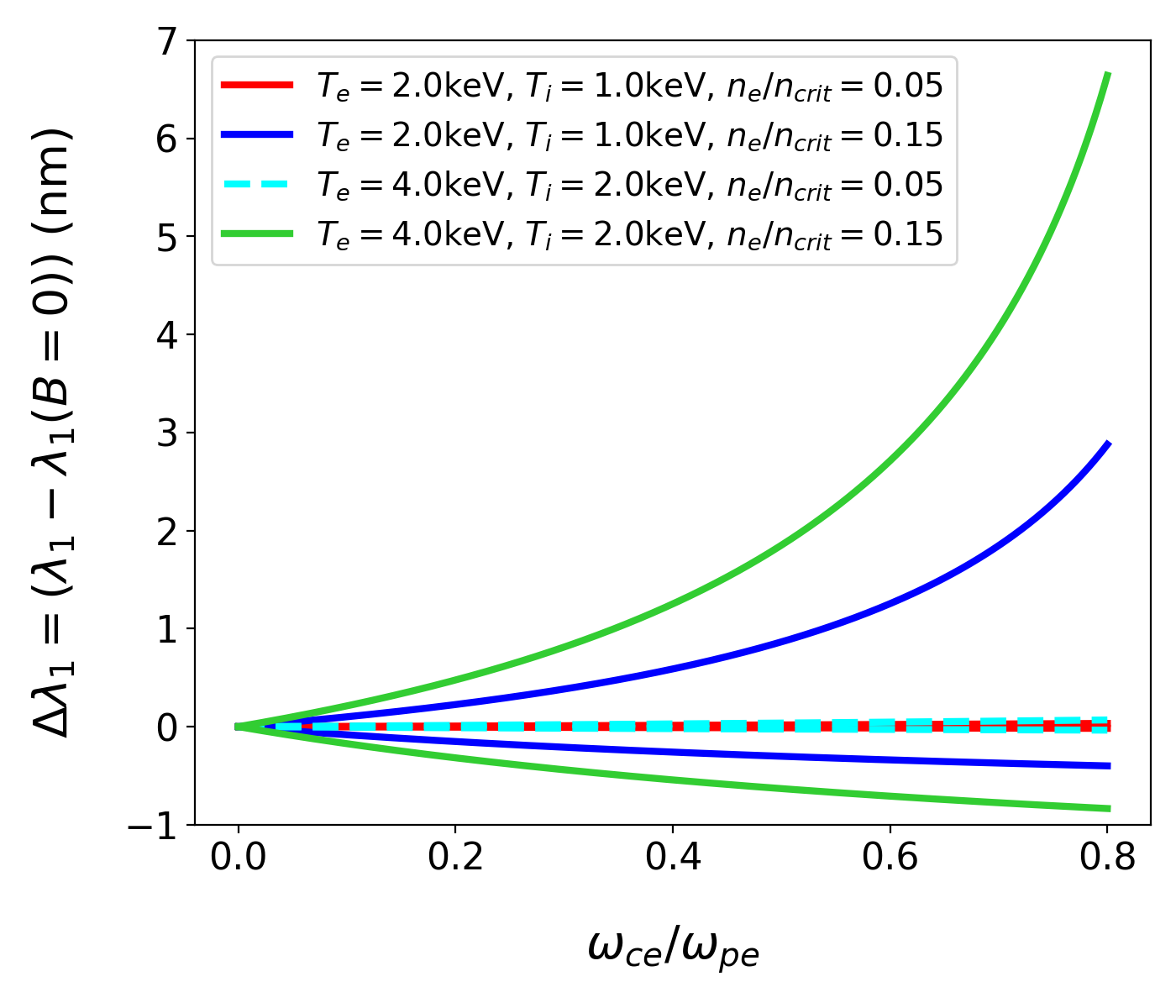}}\\
\subcaptionbox{\label{SRS_b_vs_wce}}{\includegraphics[width=8.5cm, trim={0.1cm 0.5cm 0.3cm 0.3cm}, clip]{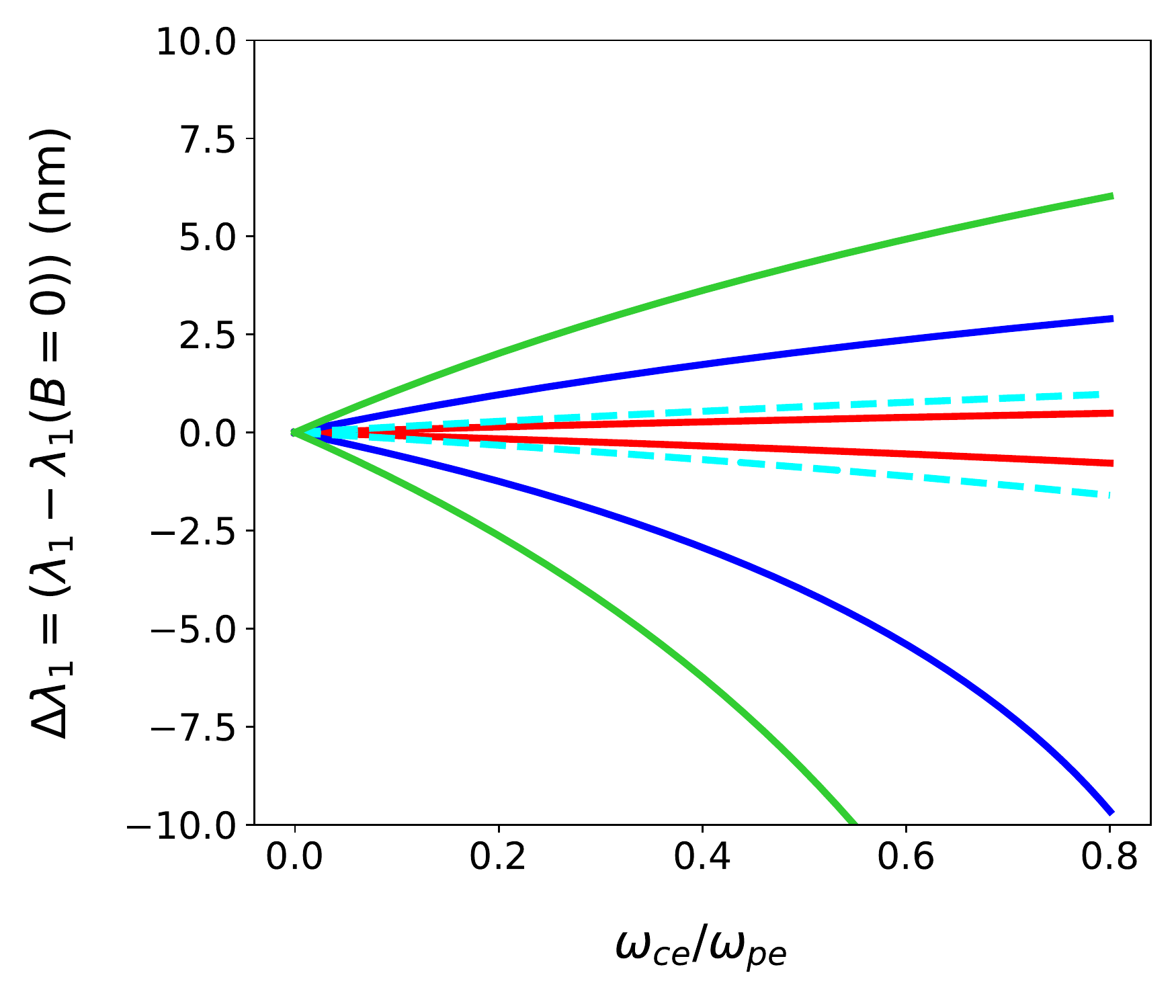}}
\caption{$\Delta \lambda_1$, the difference in wavelength of forward (\ref{SRS_f_vs_wce}) and backward (\ref{SRS_b_vs_wce}) SRS light in a magnetised versus an unmagnetised plasma, for $T_e=2.0$keV, $4.0$keV, $n_e/n_{crit} = 0.05$, $0.15$ and $\lambda_0=351$nm. For [forward, backward] SRS, $\Delta \lambda_1$ is $[>0, <0]$ for a right-polarised pump and $[<0, >0]$ for a left-polarised pump.}
\end{figure}

The shift in wavelength of SRS light due to the presence of the external magnetic field, $\Delta\lambda_1 = \lambda_1 - \lambda_{1}(\omega_{ce}=0)$, is given by
\begin{equation}
\frac{\Delta\lambda_1}{\lambda_0}=\omega_0\left(\frac{1}{\omega_1}-\frac{1}{\omega_1(\omega_{ce}=0)}\right)
\end{equation}
Substituting from equation \ref{mag_free_modes_rearr_k1}, and treating temperature and magnetic field as small perturbations in $\Omega_1$ as detailed above, we derive the following expression for $\Delta\lambda_1$ to first order in $\Omega_{ce}$ and $\Omega_{pe}^2$:
\begin{equation}
\frac{\Delta\lambda_1}{\lambda_0}\approx-\frac{\delta\Omega_1}{\Omega_{1U}^2}
\end{equation}
or equivalently
\begin{equation}
\begin{aligned}
&\Delta\lambda_1[nm] \approx -c_1\lambda^2_0[{\mu m}^2]\frac{5.48\times 10^{-4}}{\Omega_{1U}^2}Te[keV]\left(\frac{n_e}{n_{crit}}\right)^{3/2}\\
&B[T]\left(2\frac{n_{crit}}{n_e}+\sqrt{\frac{n_{crit}}{n_e}}+2\right)^{\frac{1-c_1}{2}}\sigma_0
  \end{aligned}
\end{equation}
in practical units. Under the conditions given in table \ref{tablepara}, for $n_e/n_{crit}=0.15$ and $B=100$T for SRS backscattered light from a left-polarised pump wave, the analytic approximation yields $\Delta\lambda_1=-0.041$nm, compared to the full numerical solution, which gives $\Delta\lambda_1=-0.046$nm.
Typically, in NIF-type experiments, the wavelength of back-SRS light is in the range 500-600nm, with a spectral width of 5-10nm due to damping and gradients. Given that this is the case, detecting sub-Angstrom shifts in this spectrum presents a significant challenge. 
This first-order approximation of $\Delta\lambda_1$ agrees reasonably closely with the full numerical computation of $\Delta\lambda_1$,which is plotted as a function of $\omega_{ce}/\omega_{pe}$ for $T_e=2$keV, $4$keV and $n_e/n_{crit}=0.05$, $0.15$ in figures \ref{SRS_f_vs_wce} and \ref{SRS_b_vs_wce}, for forward and back-SRS light, respectively. The effect of electron density and temperature become particularly significant for forward and backward-SRS light from a right-polarised pump as $\omega_{ce}\rightarrow\omega_{pe}$, as in this limit, $\Delta\lambda_1\rightarrow \infty$, $-\infty$, respectively.

\begin{figure}[ht!]
\subcaptionbox{\label{SRS_f_vs_Te}}{\includegraphics[width=8.5cm, trim={0.2cm 0.3cm 0.2cm 0.3cm}, clip]{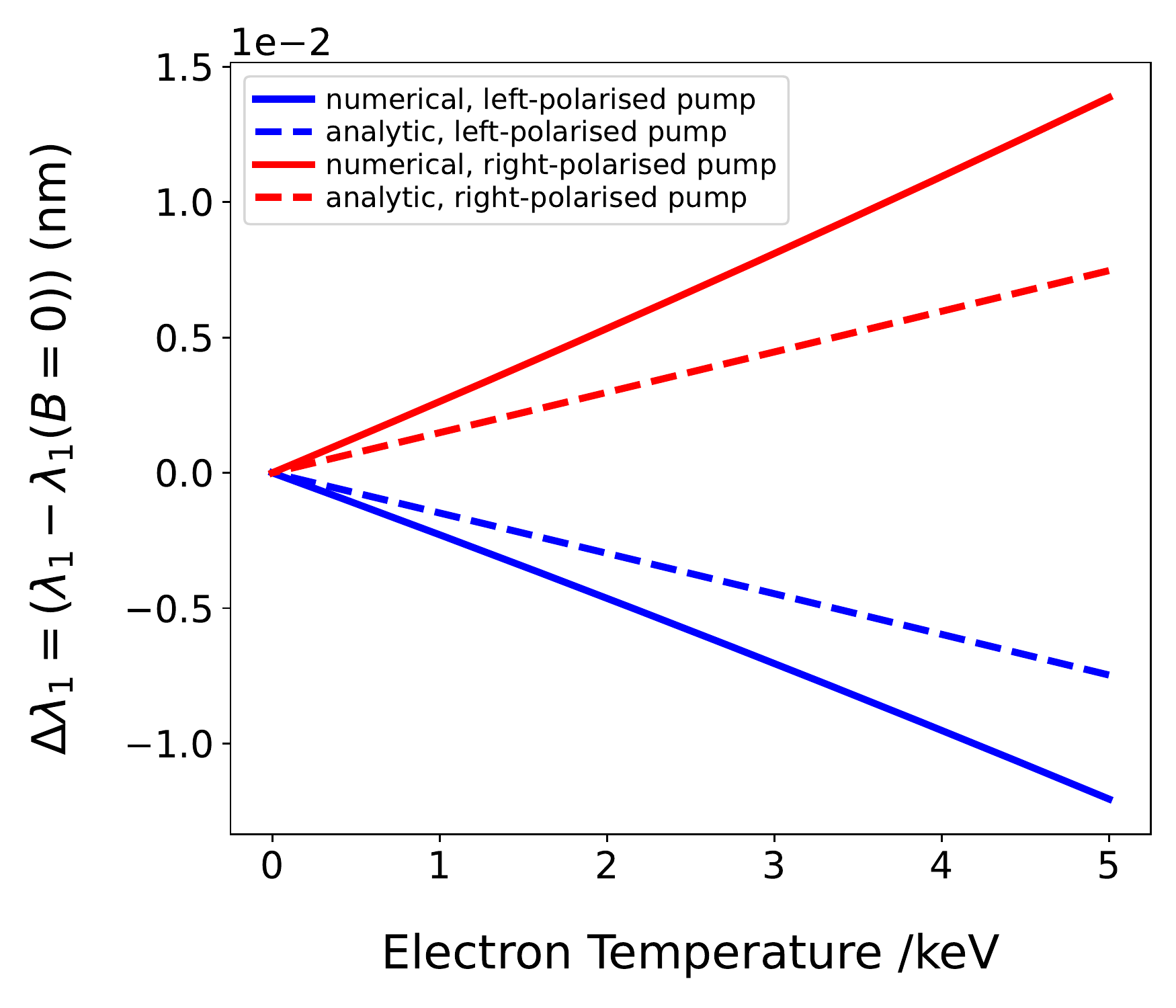}}\\
\subcaptionbox{\label{SRS_b_vs_Te}}{\includegraphics[width=8.5cm, height=7.5cm, trim={0.0cm 0.4cm 0.35cm 0.35cm}, clip]{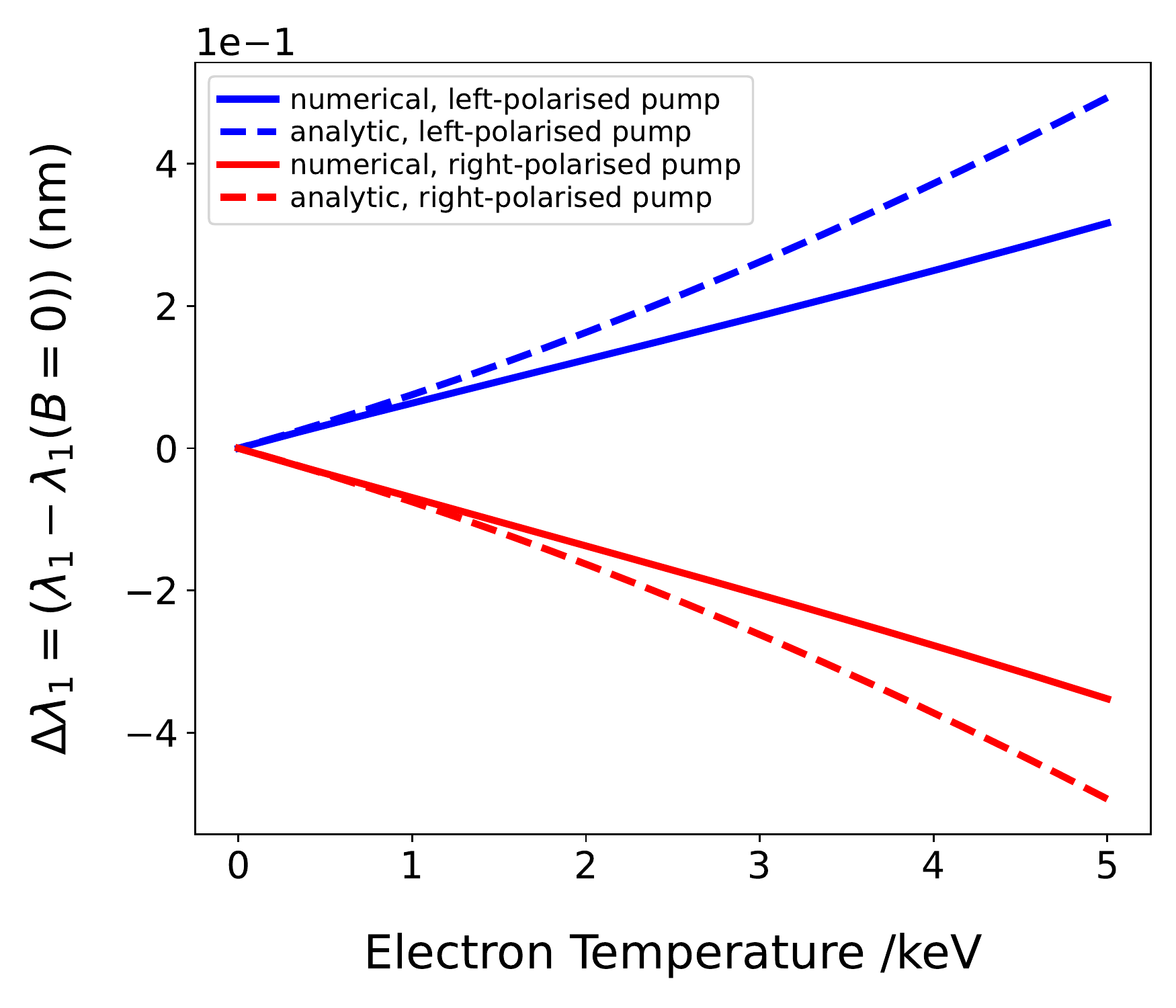}}
\caption{$\Delta \lambda_1$ of forward (\ref{SRS_f_vs_Te}) and backward (\ref{SRS_b_vs_Te}) SRS light, plotted for $\omega_{ce}/\omega_{pe}=0.1$, $\frac{n_e}{n_{crit}}=0.05$ and $\lambda_0=351$nm. Full numerical solutions are unbroken lines, first-order analytic approximations are dashed lines. \label{SRS_vs_Te}}
\end{figure}

\subsection{\label{sec:SBS}Stimulated Brillouin Scattering: SBS}

The phase matching relation for SBS, $M_B=0$ is derived in section \ref{sec:M}, and given in equation \ref{M_general}. Exact forward SBS ($c_1=1$) is not considered since in our strictly 1D geometry it does not occur.  $M_B=0$ has a spurious root for $k_2=\omega_2=0$, which connects to near-forward scatter for small but nonzero angle between $\vec k_0$ and $\vec k_1$.  The SBS growth rate is zero for $k_2=0$, so we discuss only backscatter ($c_1=-1$, $c_2=1$). For $\Omega_{ce}=0$, the exact solution is
\begin{equation}
  \Omega_{1U} = \frac{1-2\eta_0V_s + V_s^2}{1-V_s^2} \approx 1-2\eta_0V_s
\end{equation}
with $\eta_0 \equiv (1-\Omega_{pe}^2)^{1/2}$.  The approximate form for $V_s \ll 1$ is typically quite accurate.  The correction for a weak $B$ field and to leading order in $V_s^2$ is
\begin{equation}
  \delta\Omega_1 = \sigma_0\Omega_{pe}^2V_s\Omega_{ce}\left( 1+V_s\right)
\end{equation}
For simplicity, we set the final factor to 1 below.  As with SRS, the correction is ``doubly small'' since it scales with the product of $V_s \propto T_e^{1/2}$ and $\Omega_{ce}$. The scattered wavelength shift $\delta\lambda_1 \equiv \lambda_1 - \lambda_1[\Omega_{ce}=0]$, evaluated at $\Omega_{1U}=1$, is
\begin{equation}
  \frac{\delta\lambda_1}{\lambda_0} \approx -\sigma_0\Omega_{pe}^2V_s\Omega_{ce}\left( 1+V_s\right)
\end{equation}
In practical units,
\begin{equation}
\begin{aligned}
&\delta\lambda_1[Ang.] \approx -9.67\times 10^{-4}\sigma_0\frac{n_e}{n_{crit}}B[T]\\
&\sqrt{\frac{Z_iT_e[keV]}{A_i}\left(1+\frac{3T_i[keV]}{Z_iT_e[keV]}\right)}\lambda_0^2[\mu m^2]
  \end{aligned}
\end{equation}
This is an extremely small value for ICF conditions.  For the parameters shown in table \ref{table_NIF_CO2_params}, with $\lambda_0=351$ nm, $n_e/n_{crit}=0.15$, $B=100$T and a right-polarised pump, the analytic approximation gives $\delta\lambda_1\approx-2.37$pm, whereas the full numerical solution gives $\delta\lambda_1\approx-2.41$pm.

\begin{figure*}[ht!]
  \caption{Phase-matching parallelograms for backward-SBS, otherwise same as Fig.\ref{SRS_f_ne_0.15_phase_match}. Electrostatic IAW shown in black. \label{SBS_b_ne_0.15_phase_match}}
  {\begin{overpic}[width=17.0cm, trim={0.1cm 0.0cm 0.0cm 0.0cm}, clip]{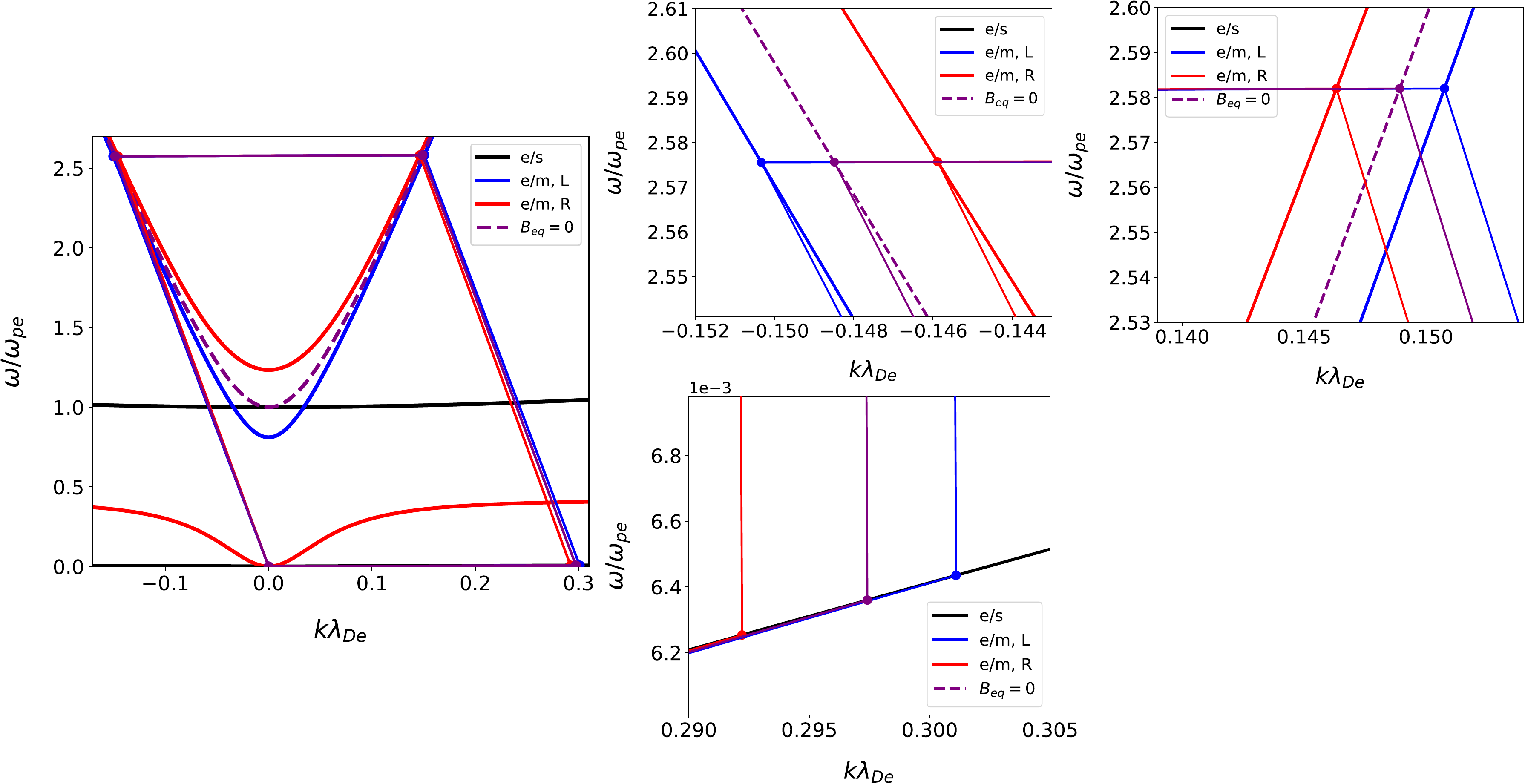}
    \put (80,40) {Pump}
\put (61,42) {Scattered}
\put (50,13) {IAW}
\end{overpic}}
\end{figure*}

%\begin{figure}[ht!]
%  \caption{[[DJS what does this figure add?  Do we need it?]] Phase-matching parallelograms for backward-SBS light, otherwise same as Fig.\ \ref{SRS_f_ne_0.15_phase_match} except $n_e/n_{crit,0}=0.05$. \label{SBS_b_ne_0.05_phase_match}}
%  \includegraphics[width=18.0cm, trim={0.1cm 0.0cm 0.0cm 0.0cm}, clip]{SBS_backward_scatter_R_L_and_0_ne_0point05.pdf}
%\end{figure}

\begin{figure}[ht!]
\centering
\begin{floatrow}
\ffigbox[\FBwidth]{\caption{$\delta\lambda_1$, the difference in wavelength of backward-SBS light in a magnetised versus an unmagnetised plasma, for three combinations of electron temperatures and densities $T_e=2.0$keV, $4.0$keV, and $n_e/n_{crit} = 0.05$, $0.15$, where the ratio of electron and ion temperature is kept constant: $T_e/T_i=2$. The laser wavelength, $\lambda_0=351$nm. The full numerical solutions and their analytic counterparts are plotted as unbroken and dashed lines, respectively. $\Delta \lambda_1$$[<0, >0]$ for a right or left-polarised pump, respectively.}\label{SBS_with_B}}{\includegraphics[width=8.5cm, trim={0.45cm 0.5cm 0.3cm 0.3cm}, clip]{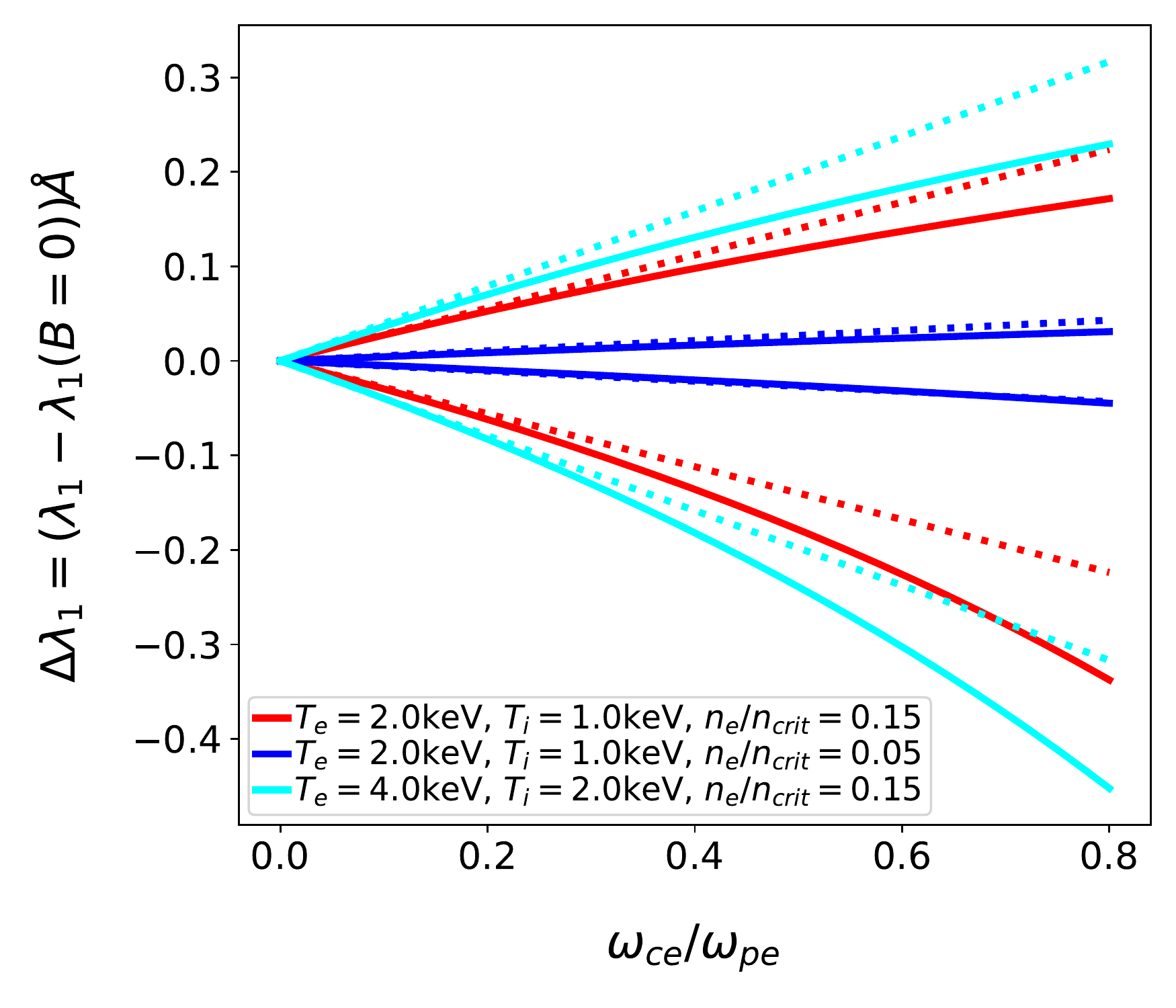}}
%{0.2\height}
\end{floatrow}
\end{figure}

\begin{figure}[ht!]
\centering
\begin{floatrow}

\ffigbox[\FBwidth]{\caption{$\Delta \lambda_1$ of backwards SBS light, plotted for $\omega_{ce}/\omega_{pe}=0.423$,  $\frac{n_e}{n_{crit}}=0.15$ and $\lambda_0=351$nm.  Full numerical solutions are unbroken lines, analytic approximations as dashed lines.\label{SBS_with_Te}}{\includegraphics[width=8.5cm, trim={0.1cm 0.5cm 0.3cm 0.4cm}, clip]{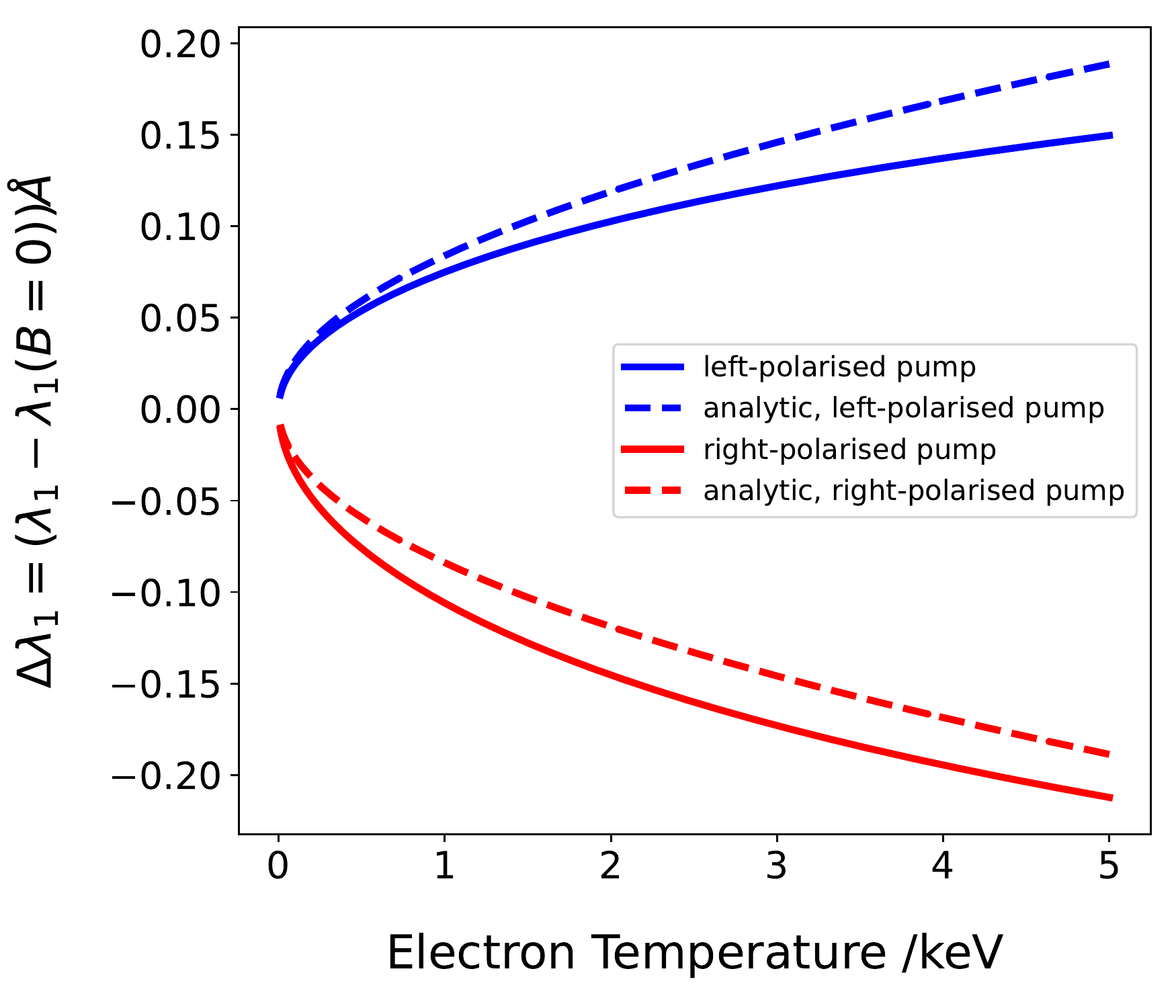}}}

\end{floatrow}
\end{figure}

\subsection{\label{sec:SWS}Stimulated Whistler Scattering: SWS}

\begin{figure}
\centering
\begin{floatrow}
\ffigbox[\FBwidth]{\caption{Phase-matching parallelogram for forward SWS: $c_1=1, c_2=-1$, where $\omega_{ce}/\omega_{pe}=0.423$ and $T_i=T_e/2$.}\label{sws_c1=1_c2=-1_phase_matching}}{\includegraphics[width=8.5cm, trim={0.6cm 0.4cm 0.4cm 0.3cm}, clip]{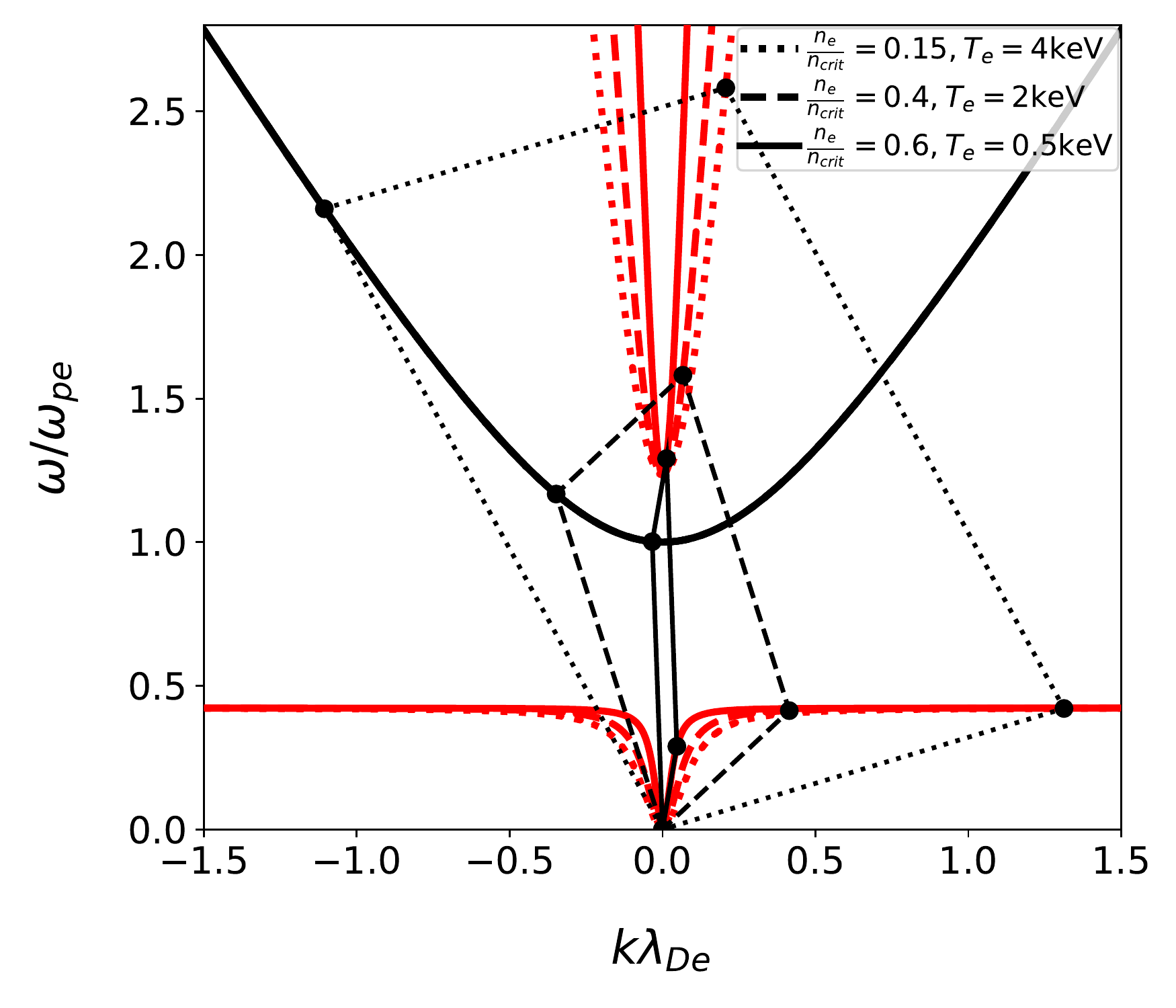}}

\end{floatrow}
\end{figure}

\begin{figure}
\centering
\begin{floatrow}

\ffigbox[\FBwidth]{\caption{Phase-matching parallelogram for backward SWS ($c_1=-1, c_2=1$), for a range of electron densities and temperatures, where $\omega_{ce}/\omega_{pe}=0.423$ and $T_i=T_e/2$.}\label{sws_c1=-1_c2=1_phase_matching}}{\includegraphics[width=8.5cm, trim={0.6cm 0.4cm 0.3cm 0.3cm}, clip]{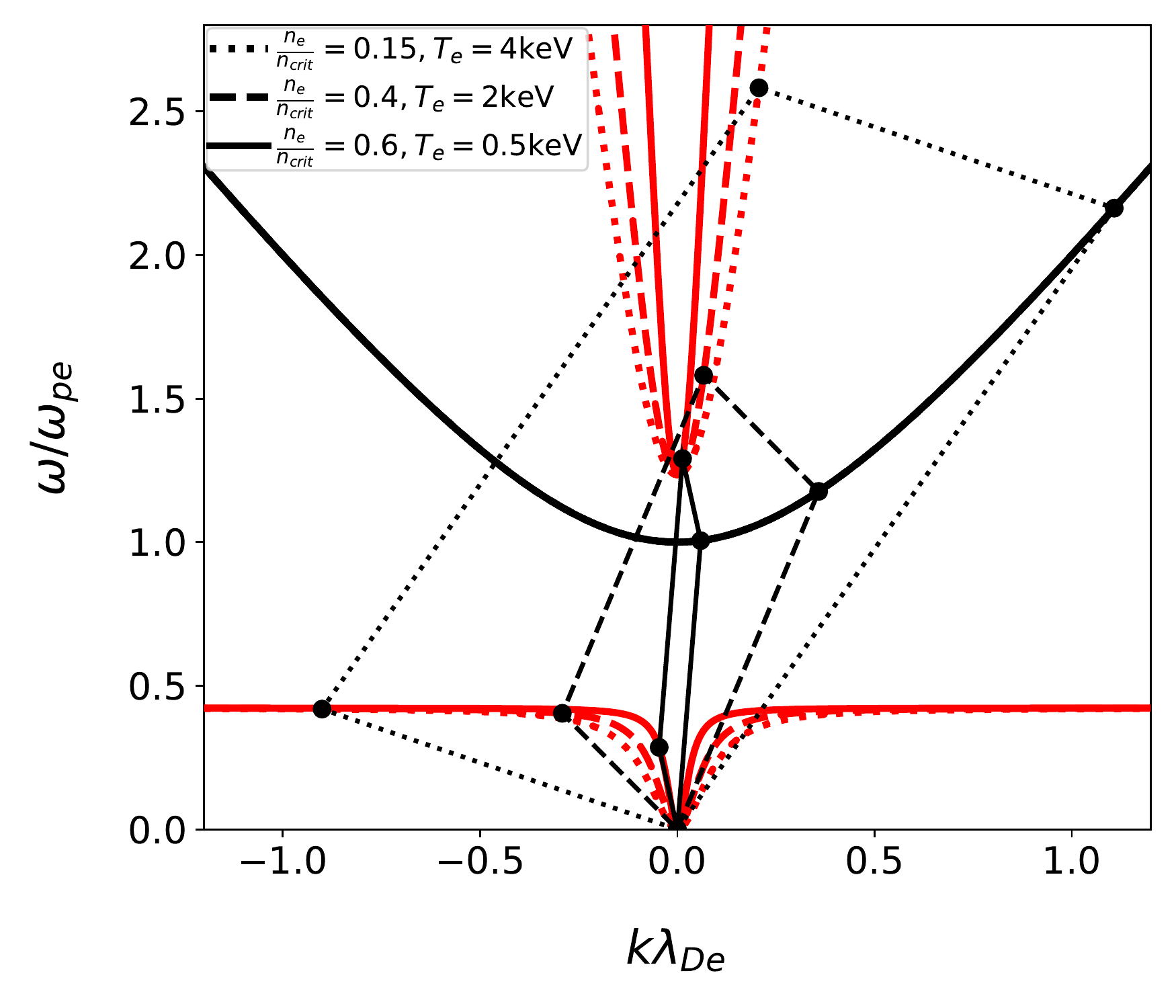}}

\end{floatrow}
\end{figure}

We now discuss SWS, which only occurs with a background magnetic field.  It resembles SRS, except the scattered e/m wave is a low-frequency whistler ($\omega_1 < \omega_{ce}$). For a cold plasma, this imposes a \textit{minimum} density of $n_e/n_{crit}\geq (1-\omega_{ce}/\omega_0)^2$ to satisfy frequency matching, as opposed to a \textit{maximum} of $n_e/n_{crit} < 1/4$ for SRS.  Forward ($c_1=+1, c_2=-1$) and backward ($c_1=-1, c_2=+1$) SWS are both kinematically allowed, though forward SWS can only occur for a plasma wave propagating counter to the pump: $c_2=-1$. The phase-matching condition $M_{RW}$ for SWS, given in equation \ref{M_general}, is identical to that of SRS except that $c_2=\pm1$.  Figures \ref{sws_c1=-1_c2=1_phase_matching} and \ref{sws_c1=1_c2=-1_phase_matching} show SWS phase matching diagrams for the allowed geometries and for a range of $n_e/n_{crit}$, $\omega_{ce}/\omega_{pe}$ and $T_e$.

\begin{figure}[ht!]
\centering
\begin{floatrow}

\ffigbox[\FBwidth]{\caption{Frequency (unbroken lines) of forward-SWS scattered light $(c_1=1, c_2=-1)$, and Langmuir wave $k_2\lambda_{De}$ (dashed lines) for various plasma densities, $n_e/n_{crit}=0.6, 0.15$, and species temperatures, $T_e=4, 0.5$keV, $T_i=T_e/2$keV. $k_2\lambda_{De}$ is plotted to indicate the strength of Landau damping.} \label{sws_w1_w0_vs_wce_c1=1_c2=-1}}{\includegraphics[width=8.5cm, trim={0.45cm 0.4cm 0.35cm 0.3cm}, clip]{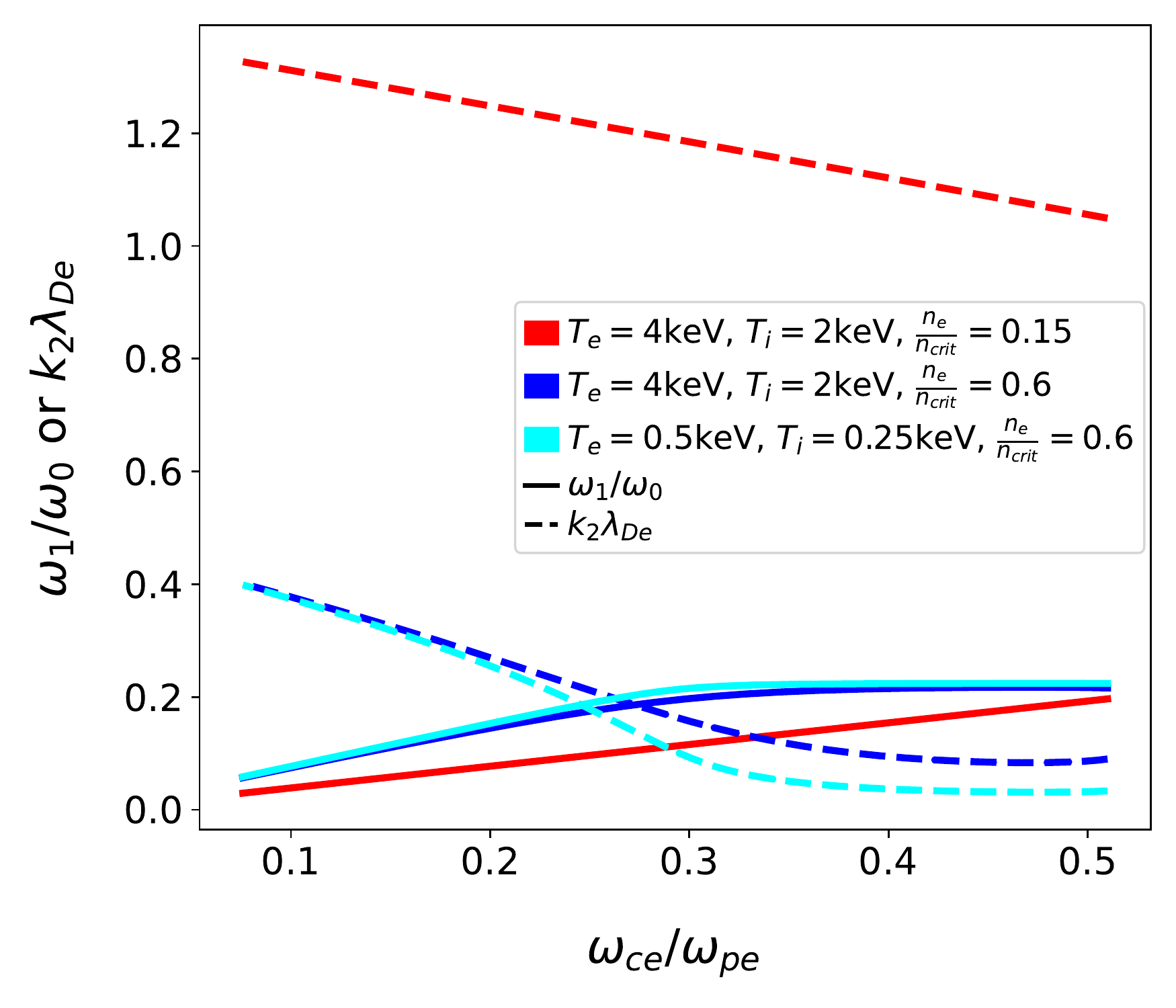}}

\end{floatrow}
\end{figure}

\begin{figure}[ht!]
\centering
\begin{floatrow}

\ffigbox[\FBwidth]{\caption{Frequency backward SWS light with $c_2=1$, for $n_e/n_{crit}=0.6, 0.15$ and $T_e=4, 0.5$keV.}\label{sws_w1_w0_vs_wce_c1=-1_c2=1}}{\includegraphics[width=8.5cm, trim={0.45cm 0.45cm 0.35cm 0.3cm}, clip]{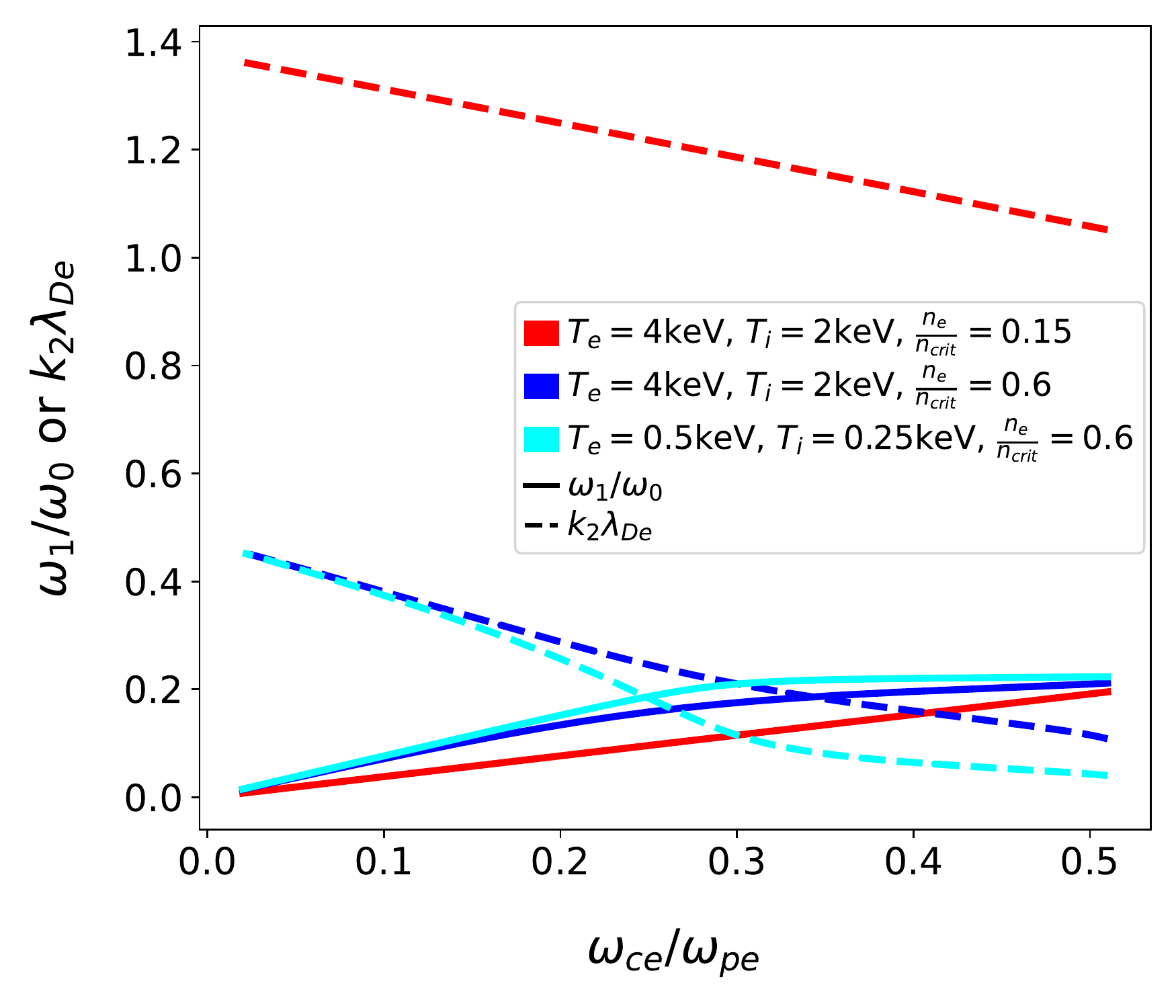}}

\end{floatrow}
\end{figure}

The relationship between $\omega_{1}/\omega_0$, $k_2\lambda_{De}$ and $\omega_{ce}/\omega_{pe}$ is shown in figures \ref{sws_w1_w0_vs_wce_c1=-1_c2=1} and \ref{sws_w1_w0_vs_wce_c1=1_c2=-1} for $(c_1, c_2)=(-1, 1), (1, -1)$, respectively, for a range of plasma densities and temperatures. The frequency of the scattered EMW increases with increasing magnetic field strength, before saturating. The rate of increase with $\omega_{ce}/\omega_{pe}$, and the values of $\omega_{1}/\omega_0$ and $\omega_{ce}/\omega_{pe}$ at which saturation occurs vary with plasma density and temperature. Increasing $T_e$ decreases the $\omega_1/\omega_0$ at which the trend saturates, while increasing $n_e/n_{crit}$ causes the observed trend to saturate at lower $\omega_1/\omega_0$ and $\omega_{ce}/\omega_{pe}$. 
$k_2\lambda_{De}$ is plotted to indicate the magnitude of Landau damping, which is expected to significantly reduce SWS growth for $k_2\lambda_{De}\gtrsim 0.5$.  In the opposite limit, the SWS growth rate approaches zero as $k_2\lambda_{De} \rightarrow 0$.

The wavelength of SWS scattered light is
\begin{equation}
\label{approx_l1_SWS}
\lambda_1[\mu m] = \frac{\omega_{ce}}{\omega_1}\frac{10709.7}{B[T]}
\end{equation}
For the bottom rows of table \ref{table_SWS_num_vals}, $n_e/n_{crit}=0.15$, $\omega_{ce}/\omega_{pe}=0.423$, and $\omega_1 \approx \omega_{ce}$.  For a pump wavelength of $0.351\mu m$, we have $B=5000$T and $\lambda_1\approx 2.14 \mu$m.  This is in the near infrared, where detectors exist but are not commonly fielded on ICF lasers.  More realistic $B$ fields will be much lower, and $\lambda_1$ much longer.

In order for SWS scattered light to be detected, it must first leave the plasma and propagate to a detector. Given the long wavelength of SWS scattered light, there is a possibility that changing plasma conditions experienced by the wave as it propagates through the plasma may cause it to become evanescent. Consider equation \ref{EM_disp_no_ions}. Rearranging for k, we obtain:
\begin{equation}
c^2k^2=\omega^2-\frac{\omega^2_{pe}}{1-\sigma\frac{\omega_{ce}}{\omega}}
\end{equation}
We see that for $\omega^2>\frac{\omega^2_{pe}}{1-\sigma\frac{\omega_{ce}}{\omega}}$, $k$ is real and the wave can propagate. If the reverse is true, $k$ is imaginary and the wave is evanescent. $\omega_{pe}$ and $\omega_{ce}$ vary in space, and generally go to zero far from the target. If $B$ tends to zero too rapidly, the dispersion relation tends to the unmagnetised one, $c^2k^2=\omega^2-\omega^2_{pe}$. In this case, if $n_e$ exceeds the critical density of the SWS scattered light wave, the wave will be reflected and will not reach the detector. However, if the magnetic field strength decreases slowly enough and/or the electron density decreases quickly enough, the wave will escape the plasma. Then $\omega_{pe}=0$ and $ck=\omega$, that is, it becomes a vacuum light wave and can propagate to the detector.

We now discuss the variation of SWS with plasma parameters. For finite $T_e$, Langmuir-wave frequency increases, an effect comparable to an increase in electron density. This enables SWS to occur at densities lower than the minimum density in a cold plasma, given in eqn \ref{sws_ncrit}. We see this in Fig.\ \ref{sws_c1=1_c2=-1_phase_matching}, where the lowest density shown, $n_e/n_{crit}=0.15$, corresponds to the highest pump frequency and a very high Langmuir-wave frequency, $\omega_2/\omega_{pe}>$2. This requires a large $k_2\lambda_{De} > 1$, which entails considerable Landau damping and therefore a low SWS growth rate. Although growth rates are beyond the scope of this paper, other work establishes that they generally are $\propto k_2^p$ (for some power $p$) when $k_2\lambda_{De}$ is small, and decrease with increasing Landau damping for large $k_2\lambda_{De}$. This means there is an effective low-density cut-off, below which SWS is kinematically allowed but strongly damped. In the opposite limit, as $n_e$ approaches $n_{crit}$ (such as $n_e/n_{crit}=0.6$ in figures \ref{sws_c1=1_c2=-1_phase_matching} and \ref{sws_w1_w0_vs_wce_c1=-1_c2=1} and table \ref{table_SWS_num_vals}), $k_2$ becomes small and Landau damping is negligible, however the growth rate of SWS also tends to 0. There is thus an intermediate range of $n_e$ in which the growth rate is optimal, and $k_2\lambda_{De}$ is moderate. The case where $n_e/n_{crit}=0.4$ and $T_e=2$ keV shown in the figures \ref{sws_c1=1_c2=-1_phase_matching}  and \ref{sws_c1=-1_c2=1_phase_matching} and table \ref{table_SWS_num_vals} typifies this regime. 

%\begin{figure}[ht!]
%\centering
%\begin{floatrow}

%\ffigbox[\FBwidth]{\caption{Frequency of light forward-scattered from an SWS wave with $c_2=+1, -1$, depending on the plasma density and species temperature (as illustrated in figure \ref{sws_continous_transition_scattered_light_sign_dependence_on_wce}) otherwise same as Fig.\ref{sws_c1=-1_c2=1_phase_matching}.} \label{sws_w1_w0_vs_wce_c1=1_c2=-1}}{\includegraphics[width=8.5cm, trim={0.4cm 0.4cm 0.4cm 0.3cm}, clip]{sws_w1_vs_wce_with_k2lDe_c1=1_c2=-1.pdf}}

%\end{floatrow}
%\end{figure}

%\begin{figure}[ht!]
%\centering
%\begin{floatrow}
%
%\ffigbox[\FBwidth]{\caption{Frequency of light backward SWS (c1=-1), with $c_2=\pm1$, for $n_e/n_{crit}=0.8$ and $T_e=0.5$keV. $k_2\lambda_{De}$ is plotted to indicate the strength of Landau damping for each set of plasma conditions and the sign of the plasma wave wavevector, which changes at $\approx \omega_{ce}/\omega_{pe}=0.37$.}\label{sws_continous_transition_scattered_light_sign_dependence_on_wce}}{\includegraphics[width=8.5cm, trim={0.05cm 0.4cm 0.4cm 0.3cm}, clip]{c1=1_c2=1_sws_w1_w0_&_k2lDe_vs_wce.pdf}}
%%{0.2\height}
%
%\end{floatrow}
%\end{figure}
\begin{table}[!ht]
\centering
\begin{ruledtabular}
\begin{tabular}{|c|c|c|c|c|c|c|c|}
$c_1$&$c_2$ & $n_e/n_{crit}$ & $T_e$ [keV] & $\omega_{1}/\omega_{0}$ & $\omega_{1}/\omega_{ce}$& $k_2 \lambda_{De}$&$(1-\frac{\omega_{ce}}{\omega_{0}})^2$\\
\hline
%1  &  1  &  0.8  &  0.5  & 0.1054  &  0.2787  &  0.002865\\
-1  &  1  &  0.6  &  0.5  & 0.2212 &  0.6752  &  0.0592&0.452\\
1  &  -1  &  0.6  &  0.5  & 0.224 & 0.6836 &  0.0336&0.452\\
\hline
%1  &  1  &  0.8  &  4  &  0.1054  &  0.2785  & 0.0081\\
%\hline
-1  &  1  &  0.6  &  4  &  0.1995 & 0.609 & 0.1502&0.452\\
1  &  -1  &  0.6  &  4  & 0.2163 & 0.6601& 0.0883&0.452\\
\hline
%1  &  1  &  0.15  &  4  &  0.1631  &  0.9955  & N/A\\
%\hline
-1  &  1  &  0.4  &  2  &  0.2557  &  0.9557  & 0.3582&0.5365\\
1  &  -1  &  0.4 &  2  & 0.2615 &  0.9776  &  0.3479&0.5365\\
\hline
-1  &  1  &  0.15  &  4  &  0.1623  &  0.9904  &  1.1074&0.6992\\
1  &  -1  &  0.15  &  4  & 0.1631  &  0.9955  &  1.106&0.6992\\
\end{tabular}
\end{ruledtabular}
\caption{\label{table_SWS_num_vals}{Frequencies of stimulated whistler scattered light for several $n_e/n_{crit}$ and $T_e$ (ion temperature, $T_i=T_e/2$), and their corresponding values of the normalised EPW wavenumber.  For all cases, $\omega_{ce}/\omega_{pe}=0.423$. The rightmost column is the minimum $n_e/n_{crit}$ for SWS to occur in a cold plasma.}}
\end{table}

\section{\label{sec:conc}Conclusion}
% free waves
We presented a warm-fluid theory for magnetized LPI, for the simple geometry of all wavevectors parallel to a uniform, background field.  The field affects the electromagnetic linear waves in a plasma, though the electrostatic waves are unaffected for our geometry.  Specifically, the right and left circular polarised e/m waves become non-degenerate, and form the natural basis, as opposed to linearly polarised waves.  This allows for Faraday rotation, which could be significant on existing ICF laser facilities for magnetic fields imposable with current technology. The field introduces two new e/m waves, the ion cyclotron and whistler wave, with no analogues in an unmagnetised plasma.

% effect of B on SRS/SBS
% for w shift numbers, use example of SRS after Eq. 97 --> |Delta lam 1| < 1 Ang; SBS before subsec C --> |Dlam1| < 0.1 Ang.
We found a parametric dispersion relation to first order in parametric coupling, Eq.\ \ref{mat_coup}, analogous to the classic 1974 work of Drake \cite{Drake74}.  We then focused on the kinematics of phase matching for three-wave interactions. Since the right and left circular polarised light waves have different $k$ vectors for the same frequency, the background field introduces a small shift in the scattered SRS and SBS frequencies compared to the unmagnetised case.  The sign of the shift depends on the pump polarization and forward vs. backward scatter.  The shift's magnitude increases with magnetic field, electron temperature, and plasma density. The wavelength shifts are $\lesssim 1$ Ang.\ for SRS, and $\lesssim 0.1$ Ang.\ for SBS, for plasma and magnetic field conditions currently accessible on lasers like NIF. Such small shifts would be extremely challenging to detect.

% SWS
The new waves supported by the background B field also allow new parametric processes, such as stimulated whistler scattering (SWS) which we studied in detail. In this process, a light wave decays to a whistler wave and Langmuir wave. This is analogous to Raman scattering, with the whistler replacing the scattered light wave. We expect SWS scattered light to be infrared, with wavelength 1 to 100 $\mu$m for fields of 10 kT to 100 T.  The whistler wavelength was found to decrease with increasing magnetic field strength, and increase with increasing plasma density and temperature.  In a cold plasma ($T_e=0$), there is a \textit{minimum} density for SWS to satisfy phase matching, namely $n_e/n_{crit} > (1-\omega_{ce}/\omega_0)^2$. Finite $T_e$ allows us to circumvent this limit, at the price of high Langmuir-wave $k\lambda_{De}$ and thus strong Landau damping.  We expect an analysis of SWS growth rates, including Landau damping, to show maximum growth for moderate $k\lambda_{De}$.

% future work
Much work remains to be done on magnetized LPI.  This paper does not discuss parametric growth rates, though they are contained in our parametric dispersion relation (without damping or kinetics), and others have studied them in the limit of weak coupling~\cite{shi19}.  It is important to know when the two circularly-polarised light waves generated by a single linearly-polarised laser (incident from vacuum) should be treated as independent pumps, with half the intensity of (and thus lower growth rates than) the original laser.  This likely occurs when the wavevector spread exceeds an effective bandwidth set by damping, inhomogeneity, or parametric coupling

Two major limitations to our model are the restriction to wavevectors parallel to the background field, and the lack of kinetic effects especially in the plasma waves.  Propagation at an angle to the B field opens up many rich possibilities, including waves of mixed e/m and e/s character, and B field effects on the e/s waves.  In the case of perpendicular propagation, the e/s waves become Bernstein waves.  Adding kinetics is essential to understanding parametric growth in many systems of practical interest, where collisionless (Landau) damping is dominant.  This also raises the so-called ``Bernstein-Landau paradox'', since Bernstein waves are na\"ively undamped for any field strength.

If these issues can be resolved, we envisage magnetized LPI modelling tools analogous to existing ones for unmagnetised LPI.  This was one of the main initial motivations for this work. For instance, linear kinetic coupling in the convective steady state and strong damping limit has been a workhorse in ICF for many years, such as for Raman and Brillouin backscatter \cite{deplete} and crossed-beam energy transfer \cite{MichaelPRL_2009}. A magnetized generalization of this needs to handle propagation at arbitrary angles to the B field, as well as arbitrary field strength.  Among other things, it must correctly recover the unmagnetised limit. A suitable linear, kinetic, magnetized dielectric function will be one of the key enablers.

% thanks
It is a pleasure to thank Y.\ Shi and B.\ I.\ Cohen for many fruitful discussions.  This work was performed under the auspices of the U.S. Department of Energy by Lawrence Livermore National Laboratory under Contract DE-AC52-07NA27344.  This document was prepared as an account of work sponsored by an agency of the United States government. Neither the United States government nor Lawrence Livermore National Security, LLC, nor any of their employees makes any warranty, expressed or implied, or assumes any legal liability or responsibility for the accuracy, completeness, or usefulness of any information, apparatus, product, or process disclosed, or represents that its use would not infringe privately owned rights. Reference herein to any specific commercial product, process, or service by trade name, trademark, manufacturer, or otherwise does not necessarily constitute or imply its endorsement, recommendation, or favoring by the United States government or Lawrence Livermore National Security, LLC. The views and opinions of authors expressed herein do not necessarily state or reflect those of the United States government or Lawrence Livermore National Security, LLC, and shall not be used for advertising or product endorsement purposes.

%\nocite{*}
\bibliography{paper_outline_DJS}% Produces the bibliography via BibTeX.

\end{document}